\documentclass[12pt,a4paper]{article}

\usepackage[english]{babel}
\usepackage{epsfig}%
\usepackage{longtable}
\usepackage{multirow}
\usepackage[numbers, authoryear, comma, longnamesfirst, sectionbib]{natbib}  
\usepackage{amssymb}
\usepackage{amsmath}
\usepackage{graphicx}
\usepackage{verbatim}
\usepackage{lscape}
\usepackage{amsthm}
\usepackage{rotate}
\usepackage{rotating}
\usepackage{pdflscape}
\usepackage{booktabs}
\usepackage{subfigure}

\usepackage{setspace}
\usepackage{placeins}
\usepackage{amsfonts}
\usepackage{booktabs}
\usepackage{footnote}	 
\usepackage{float}
\usepackage[font=small,labelfont=bf]{caption}
\usepackage{fullpage}
\usepackage{adjustbox}
\usepackage{hyperref}
\usepackage{fancyref}                                                   %
\usepackage{xcolor}    
\usepackage[flushleft]{threeparttable} 
\usepackage{natbib}  
\hypersetup{
	colorlinks,
	linkcolor={red!50!black},
	citecolor={blue!50!black},
	urlcolor={blue!80!black}
}

\usepackage{lettrine} 
\usepackage{paralist} 
\usepackage{fancyhdr} 
\usepackage{titlesec} 
\usepackage{appendix} 
\usepackage[bottom]{footmisc} 
\usepackage{array} 
\usepackage{color}    
\usepackage{colortbl} 
\usepackage{libertine}
\usepackage[libertine]{newtxmath}
\usepackage[a4paper,hmargin={0.89in,0.89in},vmargin=0.89in]{geometry}
\usepackage{epstopdf}
\usepackage{dsfont}
\usepackage{amsfonts}
\usepackage{mathtools}
\usepackage{amsmath}
\usepackage{setspace}
\usepackage{graphicx} 
\usepackage[authoryear]{natbib}
\usepackage{amssymb}
\usepackage{float}
\usepackage[english]{babel}
\usepackage{booktabs}
\usepackage{subfigure}
\usepackage{caption}
\usepackage{placeins}
\usepackage{lscape}

\theoremstyle{definition}

\begin{document}
	
	\title{Volatility jumps and the classification of monetary policy announcements}
	\author{Giampiero M. Gallo \thanks{Italian Court of Audits (Corte dei conti – disclaimer), New York University in Florence, and CRENoS}
		\and
		Demetrio Lacava\thanks{Corresponding author. University of Messina, Department of Economics. Email:dlacava@unime.it.}
		\and
		Edoardo Otranto \thanks{University of Messina, Department of Economics, and CRENoS.}}
\date{}
	\maketitle
	
	\begin{abstract}
		Central Banks interventions are frequent in response to exogenous events with direct 	implications on financial market volatility. In this paper, we introduce the Asymmetric 	Jump Multiplicative Error Model (AJM), which accounts for a specific jump component of 	volatility within an intradaily framework. Taking the Federal Reserve (Fed) as a reference, we propose a new model–based classification of monetary announcements based on their impact on the jump component of volatility. Focusing on a short window following each Fed’s communication, we isolate the impact of monetary announcements from any contamination carried by relevant events that may occur within the same announcement day.
	\end{abstract}

	\vspace{0.3cm}
	
	
	\onehalfspacing
	\newpage
	\section{Introduction}
	
	The effectiveness of monetary policy communications crucially depends on a Central bank's credibility where ``the extent to which the public believes that a shift in policy has taken place when, indeed, such a shift has actually occurred'' \citep{Cukierman:1986} needs to be built through repeated and consistent actions.\\
There is a growing attention in the literature about monetary policy announcements given their role in driving financial markets beliefs and expectations, and in particular on financial return and volatility, especially actively managing financial crises or reining in inflation.

There is a large consensus on the fact that volatility increases on announcement days: in general, if a Central Bank releases a credible announcement, new information becomes available, be it in the form of a confirmation of what markets were expecting or in the form of surprises. Either way, the announcement has an impact on market activity, with increases in the number of trades that, in turn, may translate into price changes and jumps in volatility \citep[see][for a review about volatility jumps]{Andersen:Bollerslev:Diebold:2010}.
\citet{Andersen:Bollerslev:Diebold:Vega:2003,Andersen:Bollerslev:Diebold:Vega:2007}
find evidence of significant intradaily price movements in response to macroeconomic announcements. Similarly, \citet{Johannes:2004} finds how estimated jumps in a parametric jump diffusion model for daily interest rates are associated with specific macroeconomic news events. More in general, volatility models accommodate the presence of jumps: \citet{Patton:Sheppard:2015} analyse the impact of signed jumps on volatility persistence, while other papers found a limited importance of jumps for forecasting volatility \citep{Andersen:Bollerslev:Diebold:2007,Forsberg:Ghysels:2007, Busch:Christensen:Nielsen:2011}. 

Typically a constant announcement effect is assumed on returns and volatility \citep[see, for example,][]{Wright:2012,Hattori:Schrimpf:Sushko:2016,Joyce:Lasaosa:Stevens:Tong:2011,Bomfim:2003}: we depart from such an approach by allowing for a time varying impact of announcements on volatility. We do so by exploiting the availability of ultra--high frequency based estimators of volatility \citep[see][for an exhaustive review]{McAleer:Medeiros:2008}, and the possibility of deriving significant jumps from different volatility measures by following \citet{Andersen:Bollerslev:Diebold:2007}.

In this context, one class of models that has proved itself successful in modelling volatility without resorting to logarithms and allowing for a variety of dynamic specifications is the Multiplicative Error Models stemming from the seminal paper by \citet{Engle:2002} \citep[see][for a recent survey]{Cipollini:Gallo:2022}. \citet{Cipollini:Gallo:Otranto:2021} show that MEMs are more robust to the presence of measurement errors in volatility than other volatility models. \citet{Caporin:Rossi:Santucci:2017} specify a MEM with jumps (MEM--J) where a latent process is multiplied to the conditional mean of the realized volatility, capturing extremely large realizations lying on the right tail of the distribution, especially in correspondence of financial crises. More in detail, the multiplicative jump component is generated by a mixture of independent Gamma random variables with the number of components equal to the number of jumps: the latter is a random variable as well, following a Poisson distribution. 

Here, we propose a new MEM which decouples the dynamics of expected volatility into a component for the continuous part of volatility and one for the discontinuous jump component. It belongs to the class of \textit{composite MEM} \citep{Brownlees:Cipollini:Gallo:2012,Otranto:2015}, which allows for a straightforward identification and interpretation of the jump component. The series of interest is built as intradaily volatility at thirty minute bins as an aggregation of (squared) returns observed at higher frequency, and special care is devoted to accommodating time--of--day effects. 
Focusing on a short window around each Fed's communication, our model allows for the identification of the impact of monetary policy announcements on volatility in the bin following the announcement itself. This has the important feature of zooming in to a particular time of the day, rather than relying on the daily scale which could carry some contamination from other relevant events that may occur within the same announcement day. The main goal of estimated results is thus to provide the basis for a new classification procedure of monetary announcements on the basis on their impact on volatility jumps right after the time of the announcement. We perform our analysis on several tickers, showing that our classification approach is consistent across tickers and provides useful information for both policy makers and investors about the impact of monetary announcements on the volatility of specific sectors of the market.

The proposed approach belongs to the category of model-based clustering \citep[see][for a review of clustering methods]{Maharaj:DUrso:Caiado:2019,Liao:2005}. Importantly, \citet{Aghabozorgi:Shirkhorshidi:Wah:2015} distinguish between whole, subsequent and time-point clustering ``aimed at detecting both expected and unusual patterns through the identification of dynamic changes in time series features''. Our approach belongs to the time-point techniques, not involving the classification of the full time series under investigation \citep[see][for applications on financial time series]{Caiado:Crato:2007,Otranto:2008,DeLuca:Zuccolotto:2011}.

The paper is structured as follows. Section \ref{sec:dataset} documents the construction of the dataset, where  the volatility estimators are defined, together with the identification of significant jumps (Section \ref{sec:definiton}), discusses the empirical regularities of intradaily volatility (Section \ref{sec:stylized_facts}) and the relationship between monetary policy announcements and volatility jumps (Section \ref{sec:jump_ann}). Section \ref{sec:model} introduces the model, with estimation results in Section \ref{sec:results}. The novel model--based classification approach is presented in Section \ref{sec:classification}, while Section \ref{sec:classification_results} shows the empirical results. Finally, Section \ref{sec:conclusion} concludes with some remarks.
	
	\section {An intradaily view of policy announcements}
	\label{sec:dataset}
	
	In contrast with previous contributions, we deem it interesting to adopt a strategy to investigate whether a policy announcement has a market volatility impact at or around the time of announcement itself, using intradaily data. This requires to build suitable variables tracking market activity within the day, dividing up the opening hours into elementary bins (in what follows, five minutes, but even smaller bins can be chosen) and then build larger bins (here, thirty minutes) to derive realized volatility measures.

The goal is to exploit the properties of different volatility measures to estimate a volatility jump at this thirty--minute frequency: we model the evolution of the corresponding volatility with an intradaily Multiplicative Error Model with two components, one tracking the dynamics of the continuous part of volatility and the other reacting to the significant jumps \citep[in the sense of][ as defined in Equation (\ref{eq:jump_test}) in Section \ref{sec:definiton} below]{Andersen:Bollerslev:Diebold:2007}. The intuition is that the behaviour of volatility may be perturbed by the content of an announcement, so that a complementary goal is to classify jumps on an announcement day according to given interpretable features (local maximum, local minimum, an increase or a reduction).

	\subsection{Definitions and notation} 
	\label{sec:definiton}
	We start from an elementary price series $p_{j,t}$ (in logs) recorded at regular intradaily intervals $j=0,\ldots,N$  from opening ($j=0$) to closing for day $t$, and we compute the corresponding returns: 
\begin{equation*}
	r_{j,t}=p_{j,t}-p_{j-1,t}, \qquad  j=1,\ldots,N^*.
\end{equation*}
where $N^*$ is the number of elementary bins within the day. Correspondingly, we build the realized volatility (RV hereafter) series at a lower intradaily frequency $i=1,\ldots,N$  as the square root of the realized variance obtained as the aggregation of squared returns over $M$ elementary bins ($N=N^*/M$):
\begin{equation}
	RV_{i,t}=\sqrt{\sum_{j=(i-1)M+1}^{iM} r^2_{j,t}}, \qquad i=1,\ldots,N.
	\label{eq:rv}
\end{equation}
For example, when considering elementary five--minute bins, $N^*=78$, when $M=6$ (thirty--minute realized volatility), $N=13$.\footnote{As suggested by \citet{Liu:Patton:Sheppard:2015}, considering moderate sampling frequency provides a robust to microstructure noise estimator of volatility.} 

As noticed by \citet{BarndorffNielsen:Shephard:2004,BarndorffNielsen:Shephard:2006}, the presence of volatility jumps involves a certain bias in the RV estimator; the same authors propose a jump-robust measure of volatility, the so--called  realized bipower variation (BV): 
\begin{equation}
	BV_{i,t}=\xi^{-2} \sum_{j=(i-1)M+2}^{iM} |r_{j,t}| |r_{j-1,t}|,
	\label{eq:bv}
\end{equation}
where $\xi=\sqrt{2/\pi}$ is a scale factor representing the mean of the absolute value of the standardized normally distributed jump term. This measure is used  to distinguish, in a non-parametric way, the continuous component of volatility from the jump component. The latter is obtained as a positive difference  between $RV$ and $BV$, as  $max(0,RV_{i,t}-BV_{i,t})$ \citep[see also][]{Andersen:Bollerslev:Diebold:2007}. This measure considers as jumps all the cases where  $RV_{i,t}>BV_{i,t}$, whereas we are interested only in cases where the jump modifies significantly the profile of the volatility. 
From the characteristics of $RV_{i,t}$ and $BV_{i,t}$ in what concerns jumps, \citet{Andersen:Bollerslev:Diebold:2007}  use the test statistic for \textit{significant} jumps developed by \citet{Huang:Tauchen:2005} on the basis of the distributional assumptions stated in \citet{BarndorffNielsen:Shephard:2004,BarndorffNielsen:Shephard:2006}, i.e.,\footnote{In (\ref{eq:jump_test}) the max adjustment derives from the Jansen's inequality, as shown in \citet{BarndorffNielsen:Shepard:2004b}.}
\begin{equation}
	\begin{array}{l}
		J_{i,t}=M^{-1/2}\frac{({RV}_{i,t}-{BV}_{i,t}){RV}_{i,t}^{-1}}{[(\xi^{-4}+2\xi^{-2}-5)max(1, {TQ}_{i,t} {BV}_{i,t}^{-1})]^{-1/2}},
	\end{array}\label{eq:jump_test}
\end{equation}
where  ${TQ}_{i,t}$ is the {realized tripower quarticity}, a robust estimator for the so-called {integrated quarticity}, given by:
\begin{equation}
	TQ_{i,t}=M^{-1} \eta_{4/3}^{-3} \sum_{j=(i-1)M+3}^{iM} |r_{j,t}|^{4/3} |r_{j-1,t}|^{4/3} |r_{j-2,t}|^{4/3},
\end{equation}
where $\eta_{4/3}=2^{2/3} \Gamma(7/6) \Gamma(1/2)^{-1}$. 

Based on the jump test statistic, $J_{i,t}$, we disentangle ${RV}$ into the continuous or jump-robust component $C_{i,t}$, and a pure (discontinuous) jump component $SJ_{i,t}$ \citep{Andersen:Bollerslev:Diebold:2007}:
\begin{equation}
	\begin{array}{l}
		C_{i,t}=\mathcal{I}_{[J_{i,t}\leq\Phi_q]} {RV}_{i,t} + \mathcal{I}_{[J_{i,t}>\Phi_q]}{BV}_{i,t}\\
		SJ_{i,t}=\mathcal{I}_{[J_{i,t}>\Phi_q]}({RV}_{i,t}-{BV}_{i,t}),
	\end{array}\label{eq:continuous_jump_component}
\end{equation}
where $\mathcal{I}_{[J_{i,t}\leq\Phi_q]}$ is an indicator function and $\Phi_q$ is a specific quantile of the standard normal distribution, so that a significant jumps is identified only if $J_{i,t}>\Phi_q$.
	
	\subsection{Features of intradaily volatility}
	\label{sec:stylized_facts}
	We develop our analysis on a set of intradaily RV series relative to some assets exchanged in the NYSE market, representing different sectors:
Microsoft (MSFT), as representing the Information Technology (IT) sector; Goldman Sachs (GS) and JP Morgan (JPM) for the financial sector; Johnson \& Johnson (JNJ) for the pharmaceutical sector; 3M (MMM) and Caterpillar (CAT) for the industrial sector; Home Depot, Inc. (HD) for the customer discretionary sector. The time span covers the period January 3, 2010 -- December 31, 2021; the number of observations differs series by series and is illustrated in successive Table \ref{tab:descriptive_statistics}.\footnote{The data are built starting from the tick--by--tick series of prices in the Trade and Quote (TAQ) database, which were cleaned according to the \citet{Brownlees:Gallo:2006} procedure and then sampled at five--minute intervals.}  

\begin{figure}[h!]
	\begin{center}
		\subfigure[MSFT]{\includegraphics[height=6.5cm,width=8cm]{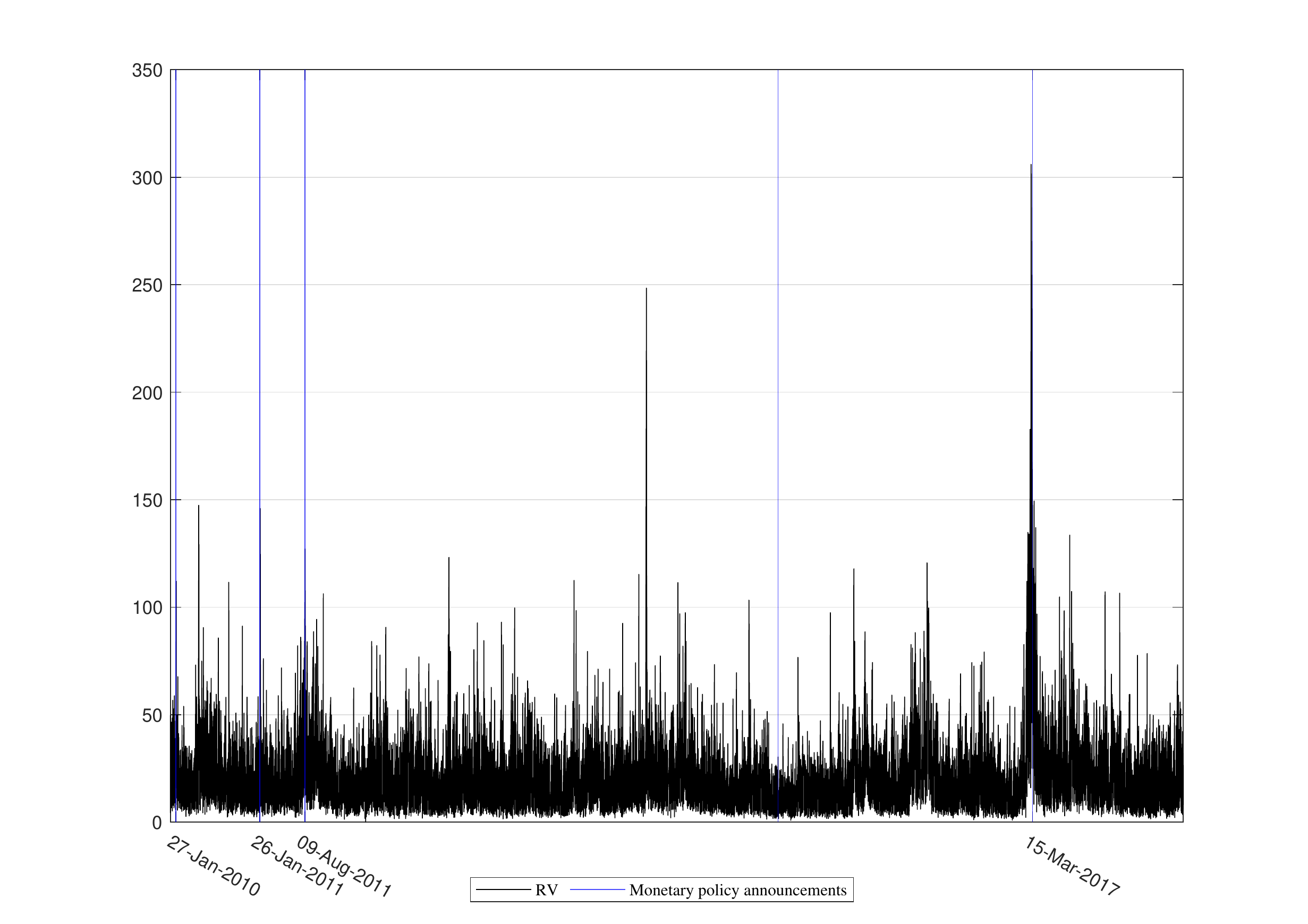}}
		\subfigure[JPM]{\includegraphics[height=6.5cm,width=8cm]{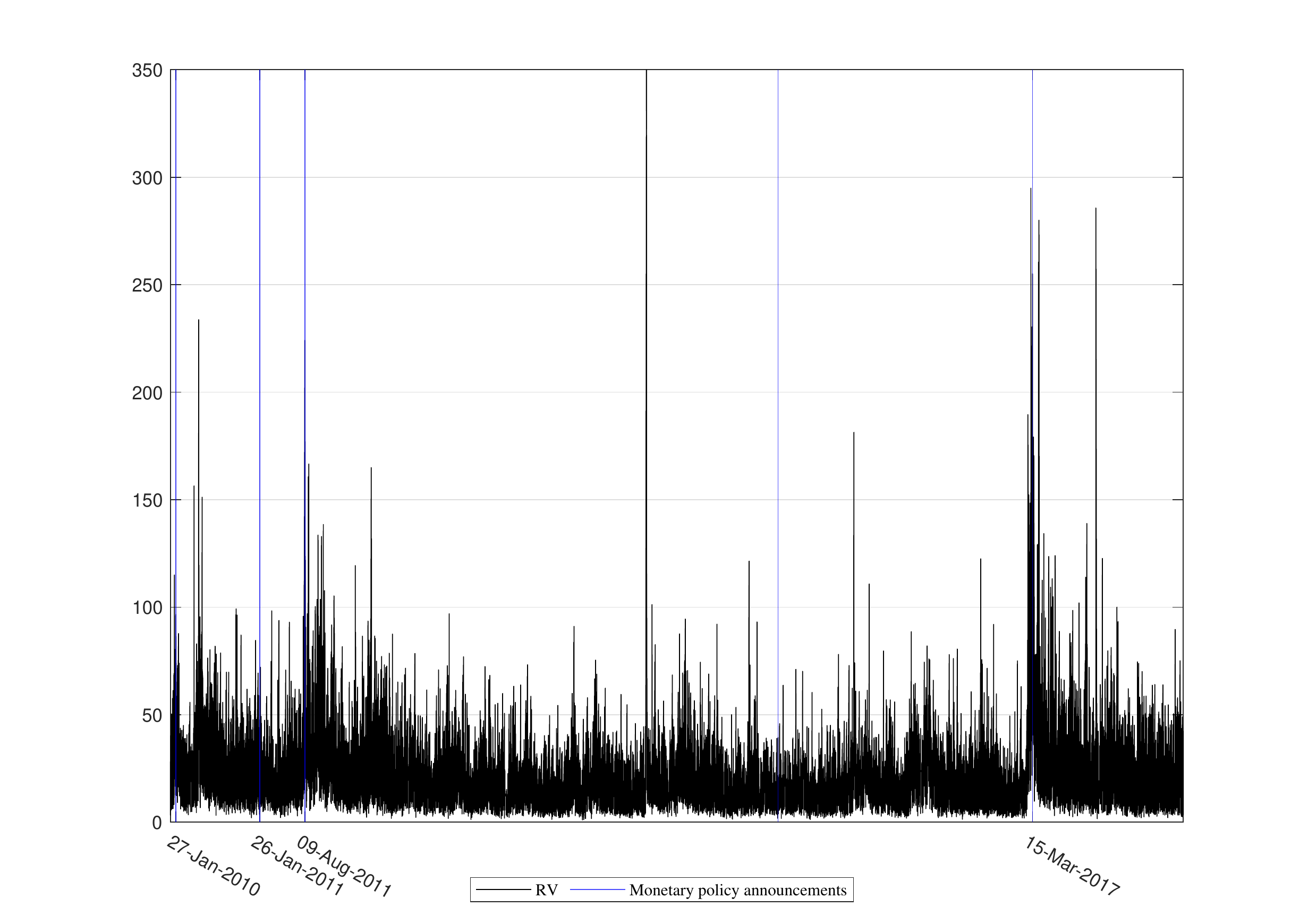}}
		\caption{30-minute realized volatility for MSFT and JPM (black line) and selected monetary policy announcements (blue line, details in the text). Sample period: January 3, 2010 -- December 31, 2021.\label{fig:rv}}
	\end{center}
\end{figure}

The evolution of the intradaily series RV according to Eq. (\ref{eq:rv}) can be shown, as an example, in Figure \ref{fig:rv} in reference to two of the series (MSFT and JPM) that we will study more in detail in Section \ref{sec:results}. A familiar pattern of persistence emerges, which is common to financial time series (even at intradaily frequency) with an occasional occurrence of bursts in the level, possibly due to the presence of jumps. Besides this commonly encountered volatility clustering, there is a changing local average volatility level. This amounts to alternating between relatively long periods of low volatility and more short-lived periods of higher volatility (for ease of reference, identifiable by the white gaps at the bottom of the series in the graph). For example, periods of low volatility are observed in the middle of the sample, while high volatility corresponds to turbulent periods for financial markets, such as the sovereign debt crisis started in the Eurozone, in the period 2010--2012, occasional flash crashes -- e.g. the ones on May 6, 2010 and August 8, 2011 -- the Brexit referendum in June 2016, and the Covid-19 pandemic starting on March 2020. Within the same figure one notices several spikes, which we investigate in relationship with the dates of the Fed's monetary policy announcements (some of which are marked as vertical blues lines).  This is the case, for example, of the announcements released on August 9, 2011 and March 3, 2020 concerning decisions about the Federal Funds Rate (FFR). Conversely, some announcements caused a spike only for some assets (e.g. the announcement on January 26, 2011 for MSFT), while in other cases one can consider the content of the announcement was anticipated by market agents, and hence had less detectable effects on volatility (cf. March 15, 2017).

\begin{figure}[h!]
	\begin{center}
		\subfigure[RV]{\includegraphics[height=6.1cm,width=8cm]{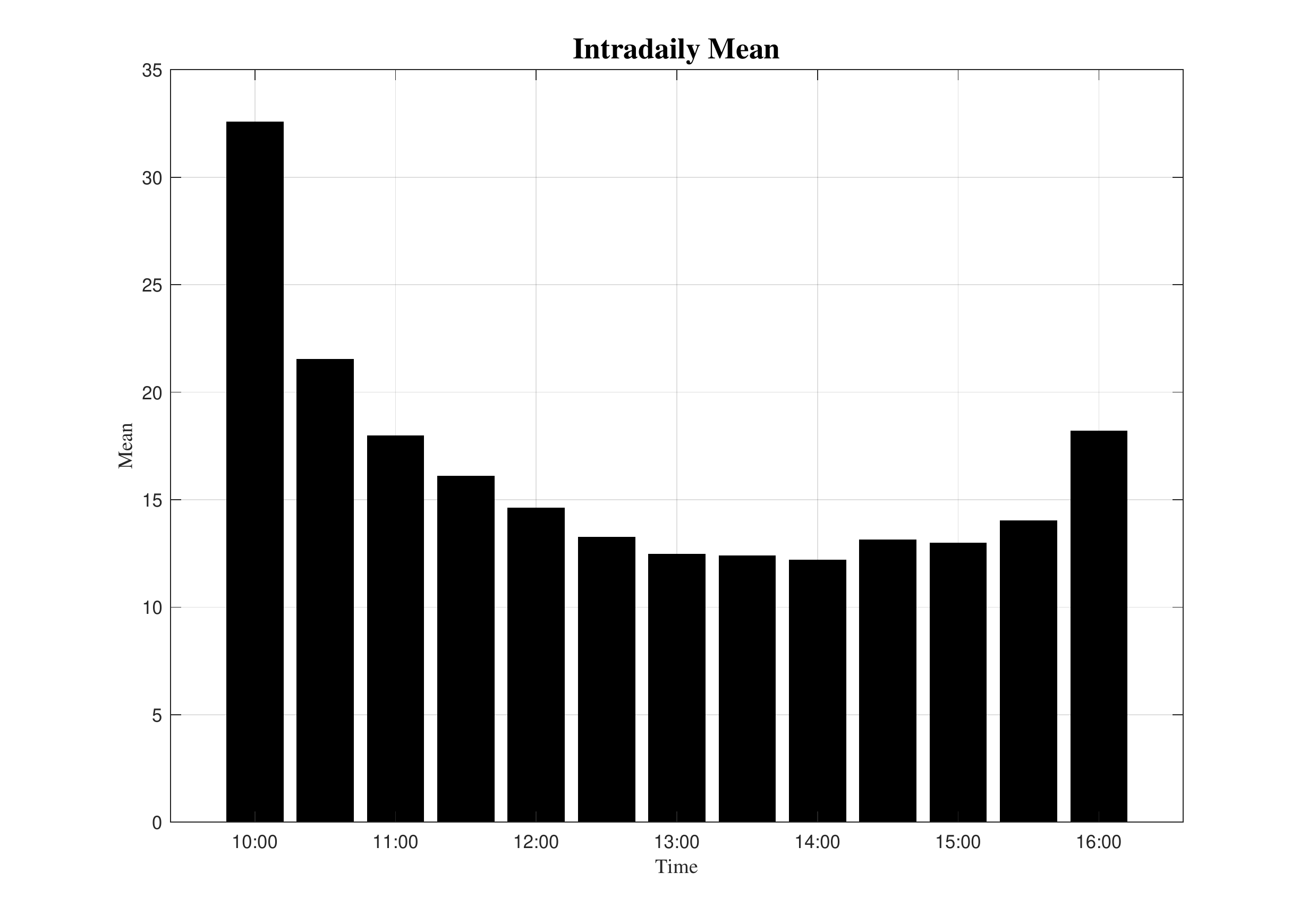}}
		\subfigure[BV]{\includegraphics[height=6.1cm,width=8cm]{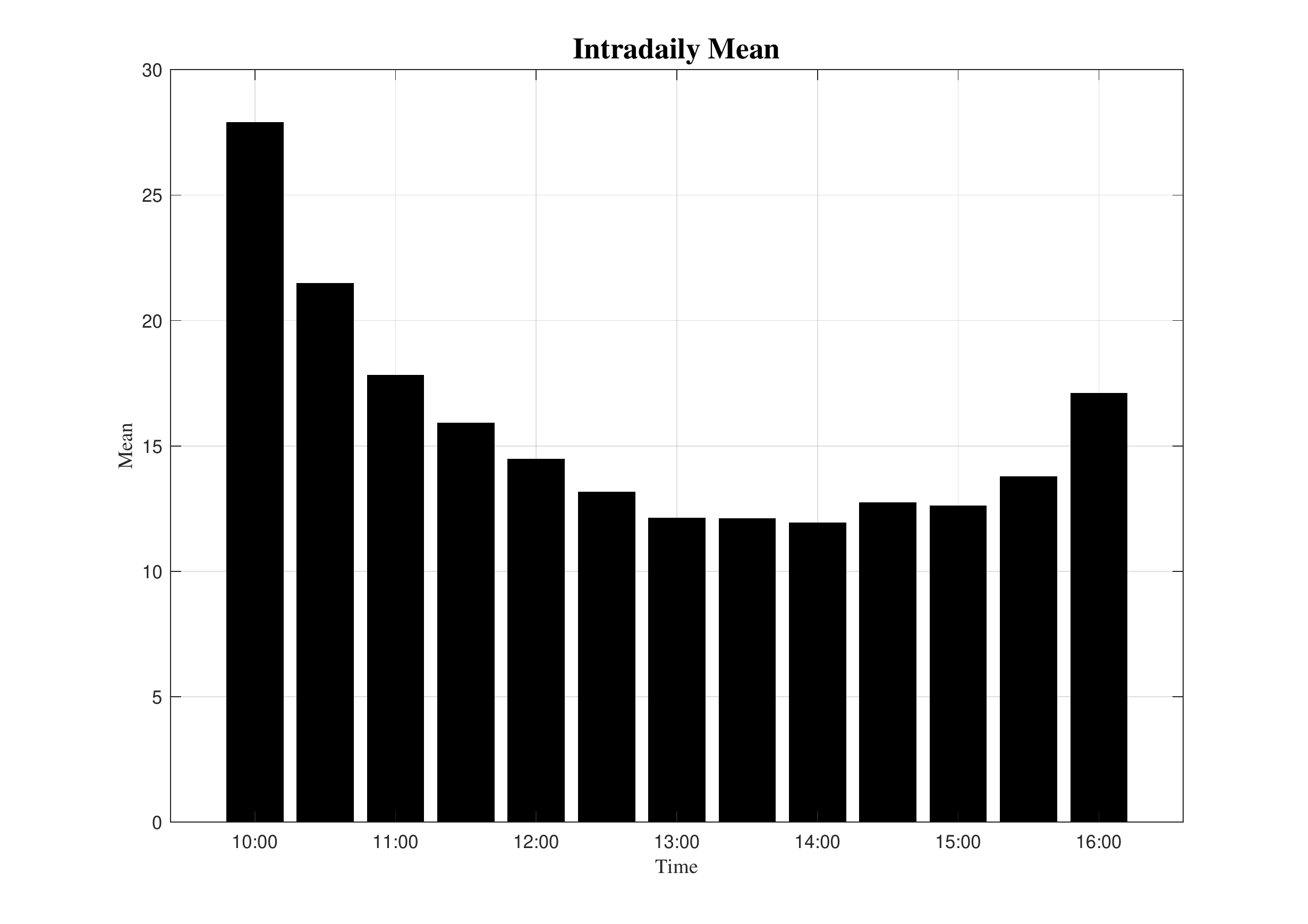}}\\
		\subfigure[RV]{\includegraphics[height=6.1cm,width=8cm]{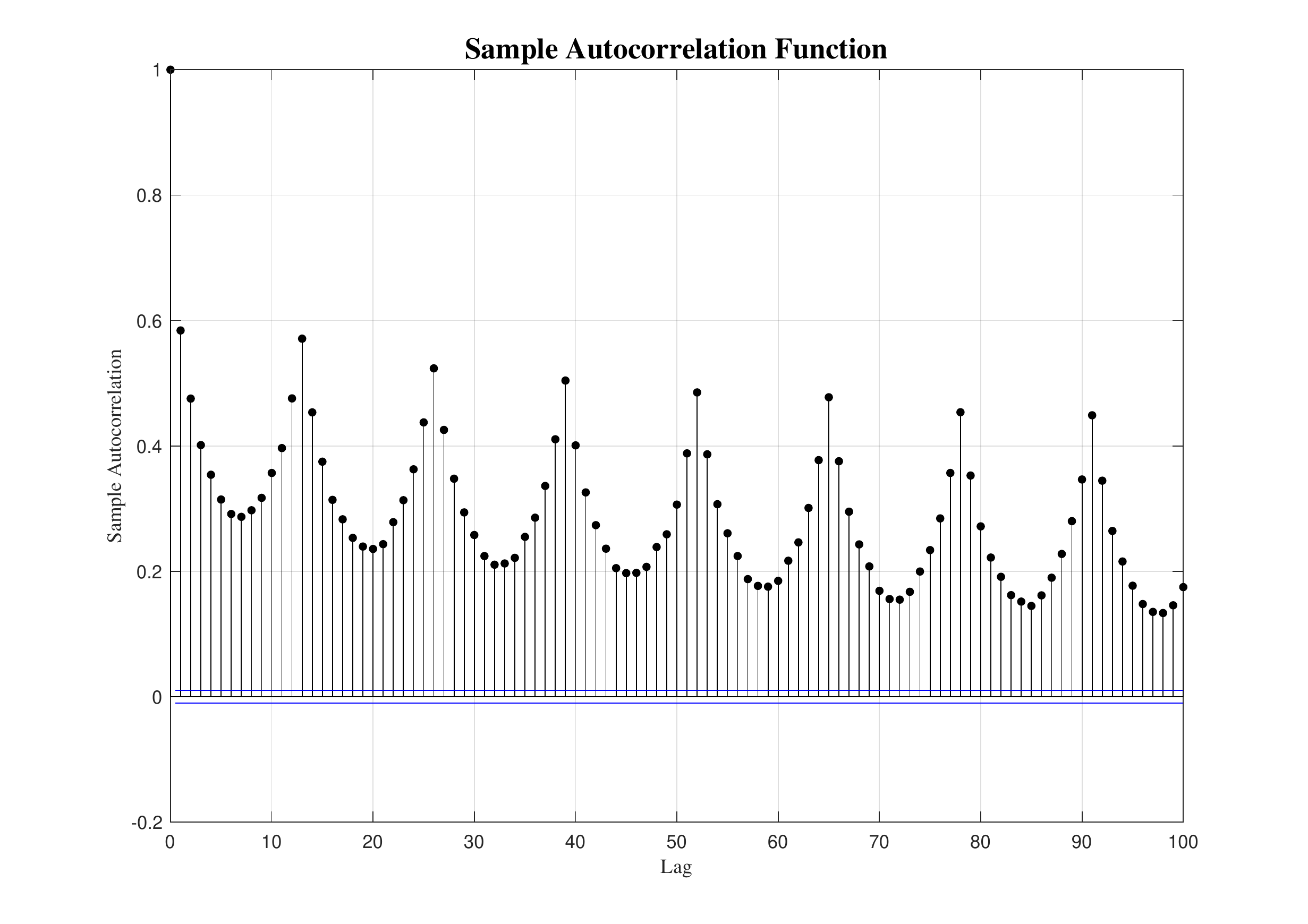}}
		\subfigure[BV]{\includegraphics[height=6.1cm,width=8cm]{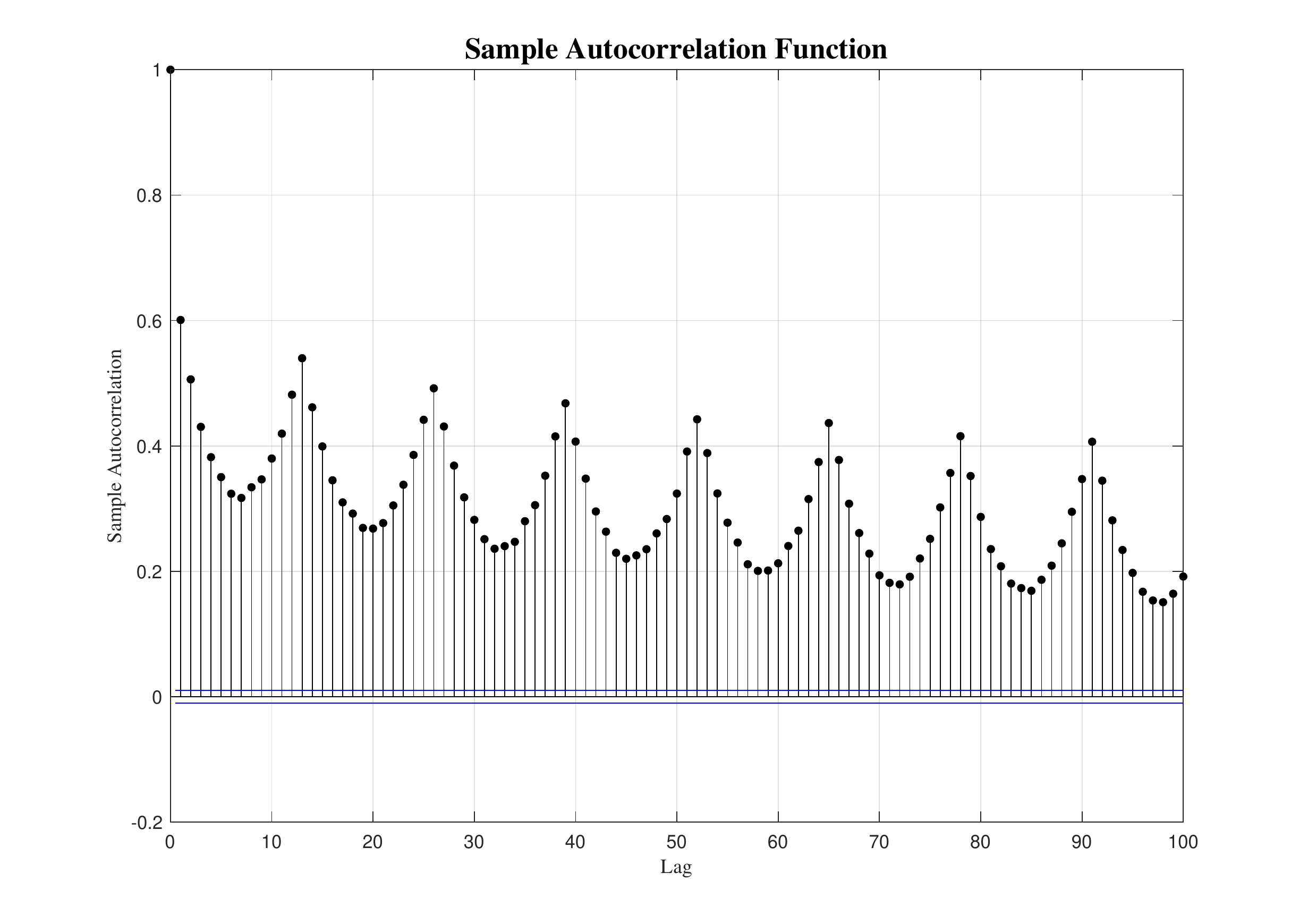}}\\
		\caption{RV and BV intradaily bin--specific means, panels (a)-(b). Sample periodogram, panels (c)-(d) for the  ${RV}$ and $BV$ of the MSFT ticker. Sample period: January 3, 2010 -- December 31, 2021.\label{fig:intraday_mean}}
	\end{center}
\end{figure}

One aspect that we need to incorporate in our subsequent analysis is that the intradaily series are characterized by time--of--day periodic effects.\footnote{This component affects even the estimation properties of the realized volatility series (BV in particular), as reported by \citet{Dette:Golosnoy:Kellermann:2022}.} In the top panels of Figure \ref{fig:intraday_mean}, for both $RV$ and $BV$ of the MSFT stock, we document a U--shaped pattern of the 13 bin--specific averages, which is common to other intradaily financial time series  \citep[e.g. financial durations as in][]{Engle:Russell:1998} and is related to the concentration of market activity at the beginning and at the end of the day.  The sample mean of the first bin of RV is 50\% (respectively, 36\% for BV) higher than the mean of the second bin, and is 143\% (112\%) higher than the mean of the rest of the day as a whole. 

The same phenomenon shows up also in the periodogram of the two realized volatility series for MSFT as in the bottom panels of Figure \ref{fig:intraday_mean}, where both RV and BV show a very persistent behaviour with regular peaks of autocorrelation.  We thus need to remove this pattern from the series in order for it not to interfere with the modelling effort: we follow the strategy of pre--filtering the $BV$ data multiplicatively, by deriving a time--of--day scale factor $\hat\pi_i$ for the series, calculated on the basis of moving averages.\footnote{Such a scale factor is calculated by the MATLAB procedure detailed as \textit{Seasonal Adjustment Using a Stable Seasonal Filter}.} 
\FloatBarrier
The adjusted BV series is therefore,
$$
\widetilde{BV}_{i,t} = BV_{i,t} \hat\pi_i,
$$
and the same periodic scale factor $\hat\pi_i$ is used to remove periodicity in RV as $$\widetilde{RV}_{i,t}=RV_{i,t}\hat\pi_i.$$
Such a strategy does not alter the original proportions between RV and BV, maintains our capability of computing volatility significant jumps, and preserves the positiveness of our variables.

\begin{figure}[h!]
	\begin{center}
		\subfigure[$\widetilde{RV}$]{\includegraphics[height=6.5cm,width=8cm]{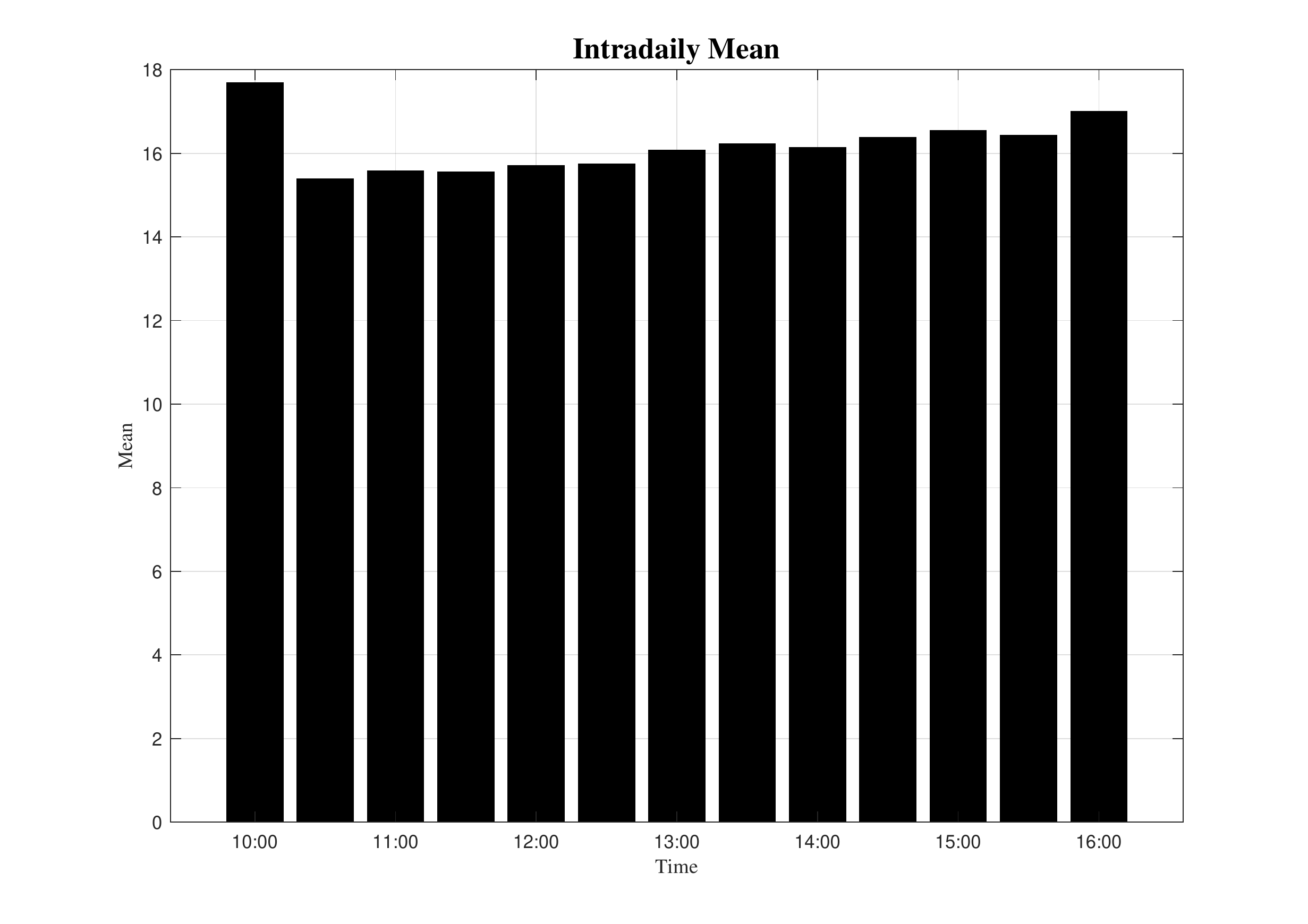}}
		\subfigure[$\widetilde{BV}$]{\includegraphics[height=6.5cm,width=8cm]{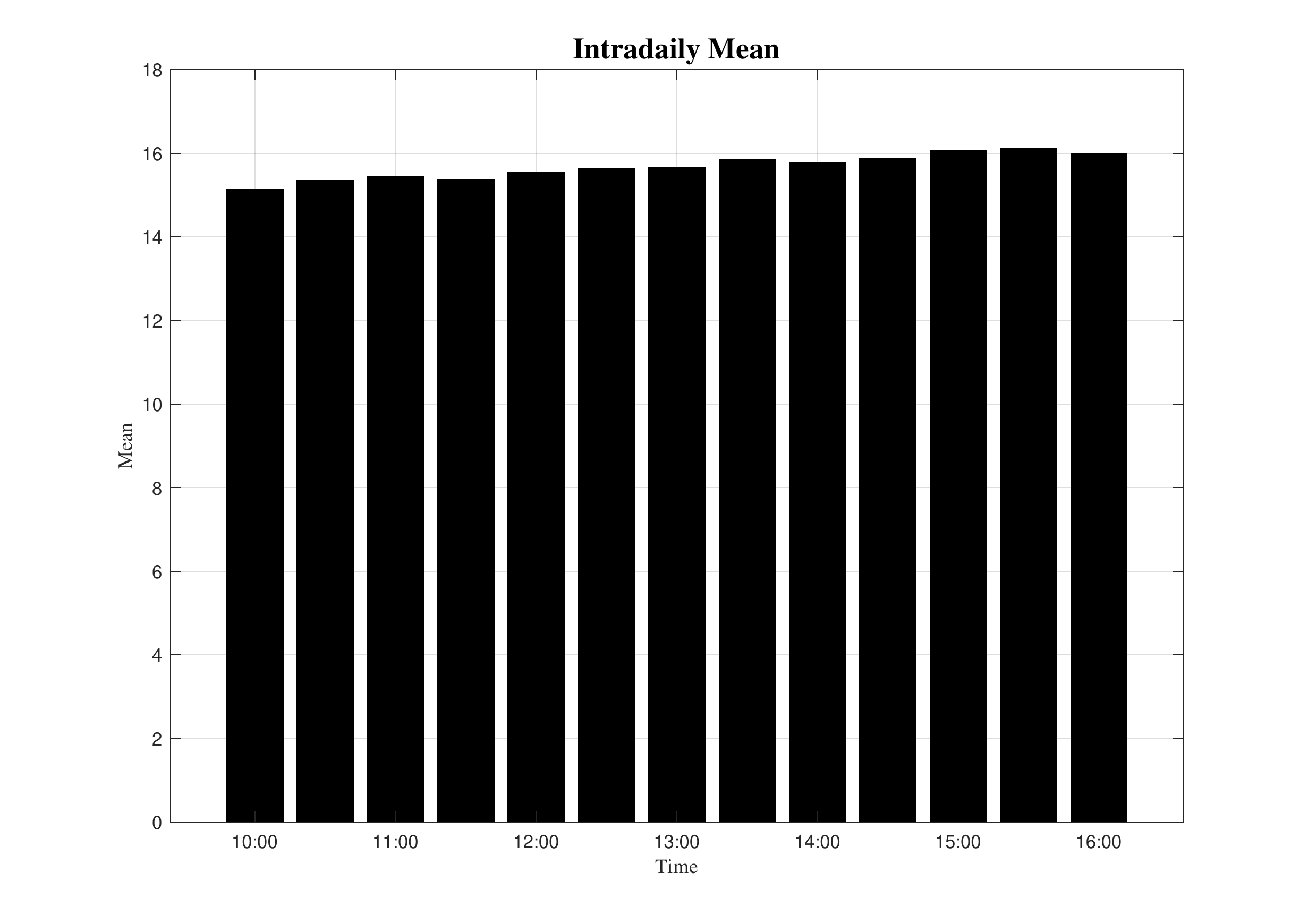}}\\
		\subfigure[$\widetilde{RV}$]{\includegraphics[height=6.5cm,width=8cm]{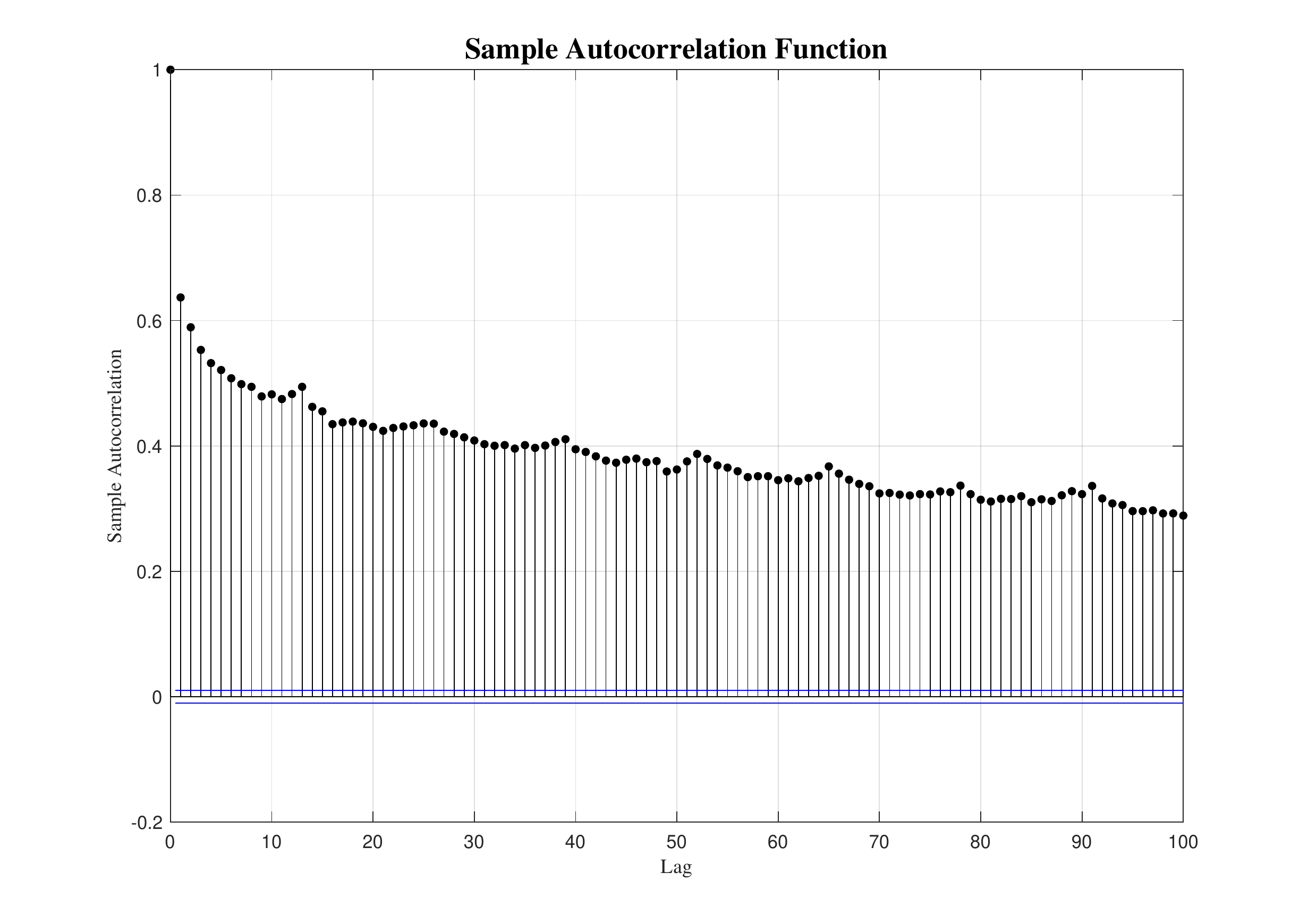}}
		\subfigure[$\widetilde{BV}$]{\includegraphics[height=6.5cm,width=8cm]{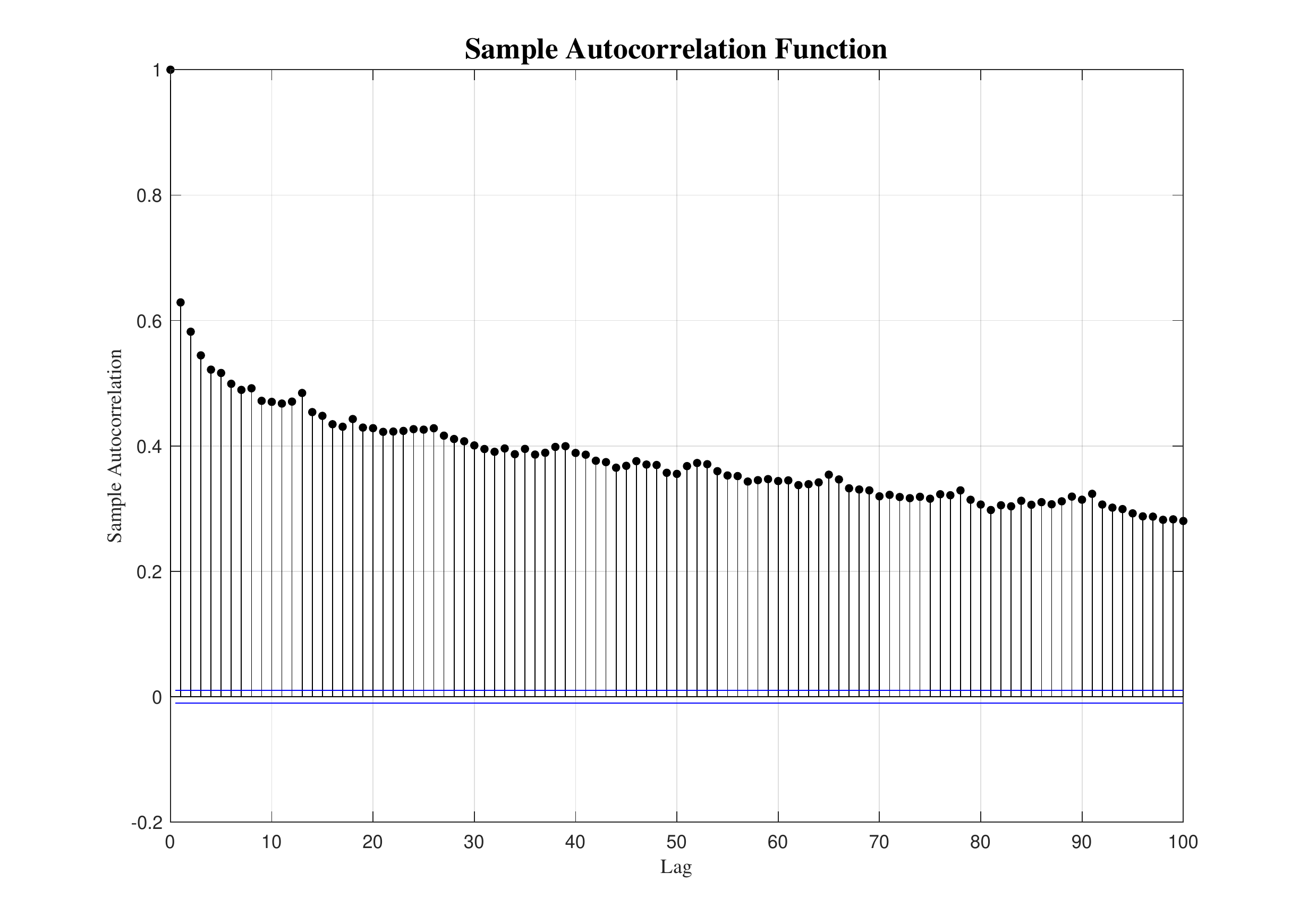}}
		\caption{RV and BV intradaily bin--specific means, panels (a)-(b). Sample periodogram, panels (c)-(d) for the  $\widetilde{RV}$ and $\widetilde{BV}$ of the MSFT ticker. Sample period: January 3, 2010 -- December 31, 2021.\label{fig:acfs}}
	\end{center}
\end{figure}

The outcome is exhibited in Figure \ref{fig:acfs}, where now the panels \textit{(a)-(b)} show the intradaily bin--specific means on the adjusted series. Interestingly, the $\widetilde{RV}$ series has an average peak in the first half--hour, which is not present in the corresponding $\widetilde{BV}$ series, a fact that is present also in other tickers with varying intensity.   The panels \textit{(c)-(d)} give the corresponding autocorrelation pattern which show that persistence is preserved and occasional peaks are much less pronounced and not at regular intervals.

A preliminary element to be considered in the subsequent modelling effort is that the first order autocorrelations are fairly stable across bins during the day (between $0.55$ and $0.7$), and similar between  $\widetilde{RV}$ and $\widetilde{BV}$: the evidence of the adjacent bins autocorrelations, namely $\rho_{h+1,h}$ with $h=1,\ldots,12$, for the two series (averaged across the seven tickers considered) is reported  in Table \ref{tab:autocorrelation}. Furthermore, Table \ref{tab:descriptive_statistics} reports the descriptive statistics of $\widetilde{RV}$ and $\widetilde{BV}$ for the selected assets, with information on the number of significant jumps reported at the bottom of the table.\footnote{We compute $SJ$ in Eq. (\ref{eq:continuous_jump_component}) choosing $q=0.55$, i.e. a value of $J_{i,t}$ bigger than $0.126$ identifies a jump as significant. We prefer to consider most jumps as significant, at our intradaily frequency, since, depending on the market activity level, even a small trade could induce a volatility jump within a trading day.} As regards the descriptive statistics, as expected, $\widetilde{RV}$ displays a higher variance and skewness than the $\widetilde{BV}$, as a consequence of the former containing more marked peaks. An average of 7 jumps per day is detected: in view of the following discussion, we note that the fraction of significant jumps recorded at 2pm, when the announcements are usually released, is about 8\%. 

\begin{table}[h!]
	\caption{Adjacent bins autocorrelations $\rho_{h+1,h}$ with $h=1,\ldots,12$ of $\widetilde{RV}$ and $\widetilde{BV}$. Sample period: January 3, 2010 -- December 31, 2021. \label{tab:autocorrelation}}
	\begin{adjustbox}{max width=0.98\linewidth,center} 
		\begin{tabular}{lcccccccccccc}
			& $\rho_{2,1}$ & $\rho_{3,2}$ & $\rho_{4,3}$ & $\rho_{5,4}$ & $\rho_{6,5}$ & $\rho_{7,6}$ & $\rho_{8,7}$ & $\rho_{9,8}$ & $\rho_{10,9}$ & $\rho_{11,10}$ & $\rho_{12,11}$ & $\rho_{13,12}$ \\
			\midrule
			$\widetilde{RV}$ & 0.556        & 0.615        & 0.654        & 0.639        & 0.637        & 0.651        & 0.694        & 0.696        & 0.563         & 0.604          & 0.681          & 0.647          \\[2mm]
			$\widetilde{BV}$ & 0.550        & 0.612        & 0.631        & 0.637        & 0.639        & 0.639        & 0.701        & 0.676        & 0.578         & 0.607          & 0.664          & 0.674     \\    
			\bottomrule
		\end{tabular}
	\end{adjustbox}
\end{table}

\begin{table}[h!]
	\caption{$\widetilde{RV}$ and $\widetilde{BV}$ descriptive statistics and number of jumps for the selected assets. Sample period: January 3, 2010 -- December 31, 2021. \textit{Note:} In parenthesis the percentage number of jumps recorded at 2pm, the usual release time of monetary policy announcements. \label{tab:descriptive_statistics}}
	\begin{adjustbox}{max width=0.98\linewidth,center} 
		\begin{tabular}{lrrrrrrr}
			& MSFT     & GS       & JPM      & JNJ      & CAT      & MMM      & HD       \\[2mm]
			\midrule
			& \multicolumn{7}{c}{$\widetilde{RV}$}                    \\[2mm]
			Mean    & 16.1903  & 18.5104  & 17.3784  & 11.8468  & 18.6788  & 13.6399  & 15.2114  \\[2mm]
			Variance& 125.1792 & 166.0786 & 162.6887 & 93.4233  & 154.3629 & 115.6457 & 125.1861 \\[2mm]
			Skewness& 4.4821   & 5.3034   & 5.6508   & 10.8788  & 3.6310    & 14.2084  & 6.3643   \\[2mm]
			Kurtosis& 50.7372  & 74.2214  & 81.9603  & 334.7588 & 33.4567  & 774.0527 & 89.8634  \\[2mm]
			\midrule
			&\multicolumn{7}{c}{$\widetilde{BV}$}                          \\[2mm]
			Mean     & 15.6820   & 17.9895  & 16.7777  & 11.4365  & 18.1290   & 13.2029  & 14.7073  \\[2mm]
			Variance & 123.7737 & 158.6057 & 153.4413 & 86.9198  & 151.3073 & 109.7661 & 120.7211 \\[2mm]
			Skewness & 4.6526   & 4.5803   & 5.0917   & 10.1127  & 3.7135   & 14.0110   & 6.2632  \\[2mm]
			Kurtosis & 57.6541  & 53.1502  & 67.4249  & 286.7568 & 35.8078  & 797.0519 & 89.3921  \\[2mm]
			\midrule
			\# of jump &19562    & 19308    & 19784    & 19709    & 19380    & 19575    & 19569   \\[2mm]
			\# of SJ &11623    & 11524    & 11997    & 11772    & 11502    & 11608    & 11672   \\[2mm]
			Total obs.   &38844    & 38961    & 38909    & 38961  & 38961    & 38883    & 38909\\
			\bottomrule
		\end{tabular}
	\end{adjustbox}
\end{table}
	
	\subsection{Jumps and announcements} \label{sec:jump_ann}
	The series of monetary announcements is obtained starting from the policy statements released by the Federal Open Market Committee (FOMC) after each meeting. Furthermore, it is possible to characterize forward guidance as a particular kind of monetary policy announcement, by exploiting the information included in the FOMC meeting minutes\footnote{Data about the date, the time and the content of monetary policy announcements are available at \url{https://www.federalreserve.gov/monetarypolicy/fomc_historical.htm}} \citep[see, for example,][]{Hattori:Schrimpf:Sushko:2016}. In the  sample period considered, 104 monetary policy announcements were released, 47 of which can be labelled as forward guidance.

\begin{figure}[tbh]
	\hspace{-.8cm}
	\includegraphics[scale=0.7]{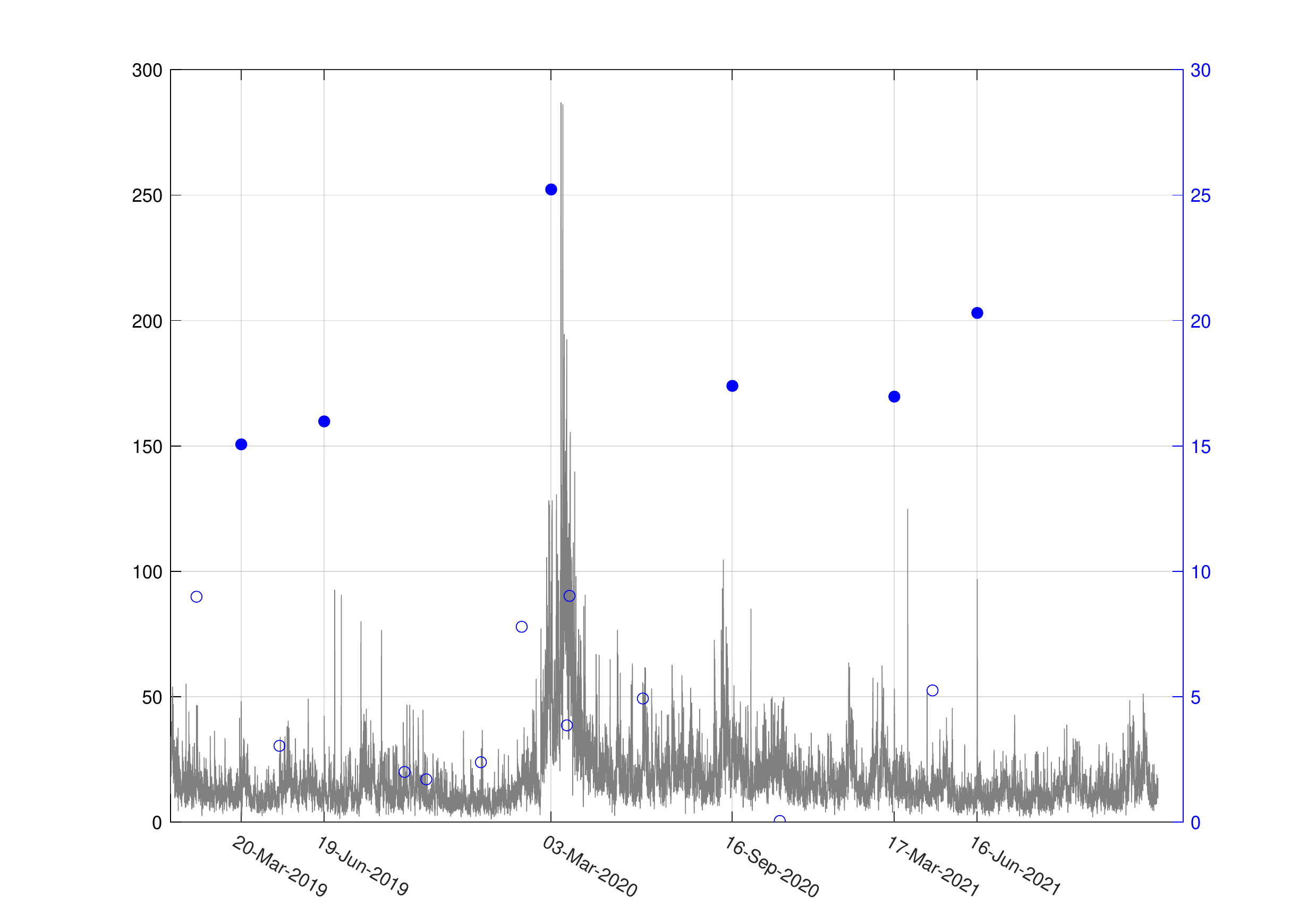}
	\caption{MSFT realized volatility grey line, measured on the left axis) and jumps (blue dots both empty and solid, on the right axis) associated to monetary announcements (details in the text). Solid blue dots identify volatility jumps accounting for more than 20\% of the overall level of volatility. Sample period: January 2, 2019 -- December 31, 2021. \label{fig:jumps_ann}}
\end{figure}

To give a visual documentation of the relationship between monetary announcements and significant volatility jumps, Figure \ref{fig:jumps_ann} shows the evolution of $\widetilde{RV}$ (grey line, left axis) together with the significant jump size observed on announcement days (blue dots -- both empty and solid -- right axis scale) between January 2, 2019 and December 31, 2021 with a total of 26 announcements, 17 of which correspond to a significant jump. On the basis of the empirical evidence in Figure \ref{fig:jumps_ann}, it is clear how the announcement effect is not constant over time: in particular, while some announcements have a limited impact on jumps, for some specific Fed's communications, significant jumps account for more than 20\% of the overall level of volatility\footnote{Noticing that $\widetilde{RV}_{i,t}=C_{i,t}+SJ_{i,t}$, the share of volatility due to significant jumps can be computed as $$\frac{SJ_{i,t}}{\widetilde{RV}_{i,t}}=1-\frac{C_{i,t}}{\widetilde{RV}_{i,t}}.$$} (blue dots). This is the case, for example, of the announcements released on March 20, 2019 and June 19, 2019, when the FOMC decided to maintain the FFR at 2.25--2.50 percent. Interestingly, a similar decision (with FFR remaining unchanged at 0-0.25 percent) on September 16, 2020, is associated to a jump marking 55\% of the total level of volatility. Yet, a significant jump is also identified on March 3, 2020, when the FFR was reduced by 0.5\%. Finally, on March 17, 2021 and June 16, 2021, the Fed decided to leave the FFR unchanged while increasing its holding of Treasury and agency mortgage-backed securities, corresponding to jumps of 35 and 20\% respectively. In synthesis, the same quantitative decision could translate into a different qualitative impact on jumps, pointing to the relevance of market expectations about the announcement rather than the decision per se.  

	\section{The model}
	\label{sec:model}

	The class of Multiplicative Error Model  \citep[MEM --][the latter allowing for asymmetric effects based on the sign of the returns -- AMEM]{Engle:2002, Engle:Gallo:2006} has proved an appropriate tool to model the volatility dynamics since it preserves the positivity of the variable without resorting to logs and it allows for a direct prediction of conditional volatility (rather than realized variance). The model takes volatility to be the product of a time-varying positive conditional mean $\mu_{i,t}$ \citep[which mirrors the GARCH-type logic,][]{Bollerslev:1986}, times an  error term $\epsilon_{i,t}$ with a positive support. 

In this paper, we propose an AMEM as the econometric model for the analysis of the impact of monetary policy announcements on volatility jumps  where the conditional mean is the sum of the two latent (continuous and jump, respectively) components of volatility. In view of the empirical regularities of intradaily volatility discussed in Section \ref{sec:stylized_facts},  in particular, a more intense market activity at the opening time of the market should be accommodated. The resulting model for $\widetilde{RV}$ is called the Asymmetric Jump Multiplicative Error Model (AJM). 
The conditional mean $\mu_{i,t}$ in the model is specified as the sum of $\varsigma_{i,t}$, i.e. the conditional expectation of the continuous part of $\widetilde{RV}_{i,t}$ (having more persistence) and the corresponding component for jumps $\kappa_{i,t}$ which reacts to the significant jumps:
\begin{equation}
	\begin{array}{l}
		\widetilde{RV}_{i,t}=\mu _{t}\epsilon _{i,t}~ \, \epsilon _{i,t}|\mathcal{F}_{i-1,t}\sim \Gamma(\vartheta ,\frac{1}{\vartheta })\\
		\mu_{i,t}=\varsigma_{i,t}+\kappa_{i,t}\\
		\varsigma_{i,t},= [\omega +\alpha_1 C_{i-1,t} +\alpha_2 C_{i-2,t}  +\beta \varsigma_{i-1,t}+ \gamma I^-_{i-1,t} C_{i-1,t}] +\delta_1 |r_{i,t^\star}|+ \delta_2 C_{i-1,t,}D_{i-1,t}  \\
		\kappa_{i,t}=\varphi \mu_{i-1,t}+\psi SJ_{i-1,t},		\\
	\end{array}\label{mod:amem_jump}
\end{equation}
where $\mathcal{F}_{i-1,t}$ is the information set at the previous bin (for the first bin of the day it is the last bin of the previous trading day). 

In the specification for $\varsigma_{i,t}$ we represent a standard AMEM(2,1) dynamics in square brackets, with a constant, two terms involving the most recent continuous part of volatility, the past conditional expectation and the asymmetric term related to the sign of the most recent return using $I^-_{i-1,t}$, as a dummy variable taking value 1 if the return of the previous bin is negative, 0 otherwise. We also need to take into consideration two important sources of additional dynamics: the first new term in the model reflects the presence of news accumulation during the night which we translate into the explicit consideration of the overnight return as a separate term $r_{i,t^\star}$.\footnote{It is given by the difference between the opening log price of day \textit{t} and the closing log price of the previous day \textit{t-1}, so that $t^\star$ denotes the time between \textit{t-1} and \textit{t}. At the end of the first bin, such information is known.} The second, separate, term reflects the need for the first bin volatility to require a particular treatment in its impact on the second bin; accordingly, we use a dummy variable $D_{i,t}$, which takes value of 1 for the first bin (30 minutes) of each trading day in our sample, i.e., $D_{i,t}=1$ for $i=1$ and $D_{i,t}=0$ for $i=2,\dots,13$. This allows us to disentangle the impact of market activity in the first bin of the day from the market activity in the rest of the day.  In this regards, it is straightforward to expect $\delta_1>0$ (the size of the overnight return, be it positive or negative, signals a surprise at opening which will have an increasing effect on volatility) and $\delta_2<0$ (less response of the second bin to the volatility recorded in the first bin).  

As for the jump component $\kappa_{i,t}$,  representing the model expectation about volatility jumps, we keep a simple specification as an AR(1) driven by the observed significant jump series with an expected positive coefficient $\psi$. 

When working with volatility components, stationarity of both $\varsigma_{i,t}$ and $\kappa_{i,t}$ are required for the model to be stationary in covariance: relying on the identifiability results of \citep{Engle:Lee:1999}, we characterize the continuous  component as being more persistent than the short lived nature of jumps, that is,  $0<\varphi<\beta<1$ in Eq. (\ref{mod:amem_jump}). As for the required positiveness, while the usual GARCH constraints ($\omega,\alpha_1,\beta,\gamma>0$) hold even in our framework, in the AJM(2,1) $\alpha_2$ can also assume negative value, with $\alpha_2>-\alpha_1\beta$ ensuring $\varsigma>0$ \citep{Cipollini:Gallo:Palandri:2020}. 

Among the distributions with positive support that can be specified for $\epsilon_{i,t}$, we opt for the Gamma distribution, because of its flexibility  (values of $\vartheta$ ensure different shapes) and in view of its robustness \citep[see][for the equivalence between first order conditions and moment conditions in the univariate case]{Cipollini:Engle:Gallo:2013}. The Gamma generally depends only on the shape parameter $\vartheta$, so that it has a unit mean and a constant variance ($1/\vartheta$): this makes the model very flexible with not only a time-varying conditional mean $E(\widetilde{RV}_{i,t}|\mathcal{F}_{i,t-1}=\mu_{i,t})$ but also a time varying conditional variance $Var(\widetilde{RV}_{i,t}|\mathcal{F}_{i,t-1}=\mu_{i,t}^2/\vartheta)$. The main implication of having a time varying volatility of volatility is that heteroscedasticity is implicit in the model, giving the opportunity of capturing possible structure in the innovations without resorting to any auxiliary regression.

Finally, Maximum Likelihood estimation involves Quasi Maximum Likelihood  (QML) properties, ensuring consistency and asymptotic normality of the coefficient estimators \citep{Engle:2002}, regardless the appropriateness of the selected distribution for the error term \citep{Engle:Gallo:2006}. However, since $\vartheta$ is unknown, robust standard errors \citep{White:1980} are a good strategy to shield against the actual shape of the  distribution.

Our model can be seen as an extension of the Composite MEM \citep[ACM,][]{Brownlees:Cipollini:Gallo:2012}, which accounts for a short and a long run component of volatility, and the Spillover AMEM \citep[SAMEM,][]{Otranto:2015}, characterized by a specific component of volatility capturing spillover effects. Furthermore, the proposed model shares some similarities with the Markov switching ACM (MS-ACM) proposed by \citep{Gallo:Lacava:Otranto:2021}, in which monetary policy announcements are classified according to their impact on a policy-related component of volatility. A major difference is that here we have a volatility component characterized by jumps rather than a regime switching specification. Moreover, our method of classifying announcements is based on the estimated jump component, which is not explicitly influenced by policy variables as in the MS--ACM.
	
	\subsection{Estimation results}
	\label{sec:results}
	
	The estimation results are presented in Table \ref{tab:estimation_results} where, for each ticker, two models are compared, differing from each other by the presence of a second lag when considering past continuous volatility.

\begin{table}[h!]
	\centering
	\caption{Estimation results with robust standard errors in parenthesis (panel a) and P--values of the Ljung-Box statistics on residuals (panel b), with $\Phi_{0.55}$. Sample period: January 03, 2010 -- December 31,2021.\label{tab:estimation_results}}
	\begin{adjustbox}{max width=1\linewidth,center}
		\begin{tabular}{lrr|rr|rr|rr|rr|rr|rr}
			\multicolumn{1}{c}{Panel {a)}} & \multicolumn{14}{c}{Estimation results}\\[2mm]
			\hline
			& \multicolumn{2}{c}{MSFT}    & \multicolumn{2}{c}{GS} & \multicolumn{2}{c}{JMP} & \multicolumn{2}{c}{JNJ} & \multicolumn{2}{c}{CAT} & \multicolumn{2}{c}{MMM} & \multicolumn{2}{c}{HD} \\
			$\omega$    & 0.5749   & 0.1173   & 0.5765   & 0.2064   & 0.5168   & 0.1141   & 0.4052   & 0.1679   & 0.5436   & 0.1115   & 0.3309   & 0.2215   & 0.4578   & 0.1803   \\
			& (0.0364) & (0.0358) & (0.0460) & (0.0399) & (0.0418) & (0.0246) & (0.0340) & (0.0246) & (0.0381) & (0.0317) & (0.0313) & (0.0389) & (0.0379) & (0.0435) \\[2mm]
			$\alpha_1$  & 0.2199   & 0.2589   & 0.1997   & 0.2463   & 0.2116   & 0.2662   & 0.2012   & 0.2442   & 0.2060   & 0.2523   & 0.1829   & 0.2229   & 0.1829   & 0.2255   \\
			& (0.0068) & (0.0061) & (0.0069) & (0.0058) & (0.0074) & (0.0061) & (0.0081) & (0.0067) & (0.0068) & (0.0059) & (0.0073) & (0.0080) & (0.0071) & (0.0068) \\[2mm]
			$\alpha_2$  &          & -0.2045  &          & -0.1706  &          & -0.2065  &          & -0.1556  &          & -0.2015  &          & -0.1081  &          & -0.1498  \\
			&          & (0.0128) &          & (0.0118) &          & (0.0098) &          & (0.0105) &          & (0.0107) &          & (0.0188) &          & (0.0167) \\[2mm]
			$\beta$     & 0.7668   & 0.8926   & 0.7905   & 0.8763   & 0.7769   & 0.8961   & 0.7956   & 0.8723   & 0.7829   & 0.9018   & 0.8032   & 0.8465   & 0.8037   & 0.8728   \\
			& (0.0088) & (0.0177) & (0.0085) & (0.0110) & (0.0108) & (0.0110) & (0.0099) & (0.0094) & (0.0093) & (0.0146) & (0.0091) & (0.0118) & (0.0093) & (0.0141) \\[2mm]
			$\gamma$    & 0.0360   & 0.0105   & 0.0268   & 0.0137   & 0.0306   & 0.0112   & 0.0155   & 0.0122   & 0.0292   & 0.0100   & 0.0348   & 0.0273   & 0.0340   & 0.0174   \\
			& (0.0033) & (0.0032) & (0.0030) & (0.0023) & (0.0031) & (0.0022) & (0.0034) & (0.0023) & (0.0032) & (0.0025) & (0.0036) & (0.0040) & (0.0032) & (0.0035) \\[2mm]
			$\delta_1$  & 0.0237   & 0.0081   & 0.0248   & 0.0137   & 0.0228   & 0.0089   & 0.0274   & 0.0154   & 0.0242   & 0.0086   & 0.0236   & 0.0173   & 0.0207   & 0.0107   \\
			& (0.0020) & (0.0023) & (0.0019) & (0.0020) & (0.0020) & (0.0017) & (0.0034) & (0.0030) & (0.0018) & (0.0022) & (0.0021) & (0.0026) & (0.0022) & (0.0023) \\[2mm]
			$\delta_2$  & -0.0614  & -0.0304  & -0.0471  & -0.0358  & -0.0496  & -0.0270  & -0.0606  & -0.0494  & -0.0467  & -0.0207  & -0.0460  & -0.0457  & -0.0472  & -0.0344  \\
			& (0.0074) & (0.0086) & (0.0076) & (0.0072) & (0.0081) & (0.0067) & (0.0083) & (0.0077) & (0.0073) & (0.0071) & (0.0077) & (0.0076) & (0.0076) & (0.0077) \\[2mm]
			$\varphi$   & -0.2136  & 0.3509   & -0.2144  & 0.2040   & -0.1952  & 0.2762   & -0.2367  & 0.1141   & -0.1968  & 0.3342   & -0.1876  & 0.0166   & -0.2147  & 0.2117   \\
			& (0.0290) & (0.0575) & (0.0312) & (0.0501) & (0.0380) & (0.0427) & (0.0392) & (0.0475) & (0.0392) & (0.0574) & (0.0388) & (0.0685) & (0.0439) & (0.0695) \\[2mm]
			$\psi$      & 0.2322   & 0.2622   & 0.2084   & 0.2476   & 0.1897   & 0.2323   & 0.1902   & 0.2368   & 0.2112   & 0.2549   & 0.2251   & 0.2579   & 0.1962   & 0.2504   \\
			& (0.0168) & (0.0151) & (0.0183) & (0.0154) & (0.0185) & (0.0143) & (0.0193) & (0.0153) & (0.0186) & (0.0145) & (0.0189) & (0.0180) & (0.0217) & (0.0169) \\[2mm]
			$\vartheta$ & 5.6346   & 5.6748   & 5.8286   & 5.8705   & 5.8098   & 5.8783   & 5.3194   & 5.3593   & 5.9627   & 6.0183   & 5.7086   & 5.7287   & 5.7592   & 5.7922   \\
			& (0.0510) & (0.0517) & (0.0551) & (0.0569) & (0.0579) & (0.0575) & (0.0589) & (0.0607) & (0.0535) & (0.0534) & (0.0668) & (0.0681) & (0.0549) & (0.0554)\\[2mm]
			\hline
			Loglik &  -121078      & -120932  & -126257  & -126109  & -123053  & -122812  & -110074  & -109921  & -125976  & -125785  & -113873  & -113801  & -118499  & -118382   \\ [2mm]
			\hline
			Panel {b)}&	\multicolumn{14}{c}{Diagnostic: Ljung-Box (p-value)}\\[2mm]
			LB1             & 0.0000       & 0.7314   & 0.0000   & 0.0005   & 0.0000   & 0.0176   & 0.0000   & 0.0994   & 0.0000   & 0.1319   & 0.0000   & 0.0000   & 0.0000   & 0.0023   \\[2mm]
			LB5             & 0.0000       & 0.1392   & 0.0000   & 0.0000   & 0.0000   & 0.0096   & 0.0000   & 0.0009   & 0.0000   & 0.0070   & 0.0000   & 0.0000   & 0.0000   & 0.0000   \\[2mm]
			LB10            & 0.0000       & 0.0100   & 0.0000   & 0.0000   & 0.0000   & 0.0017   & 0.0000   & 0.0000   & 0.0000   & 0.0108   & 0.0000   & 0.0000   & 0.0000   & 0.0000 \\[2mm]
			
			\hline
		\end{tabular}
	\end{adjustbox}
\end{table}

Focusing on the unrestricted version of our model (second column, with $\alpha_2$, negative as expected), coefficients are highly significant with a persistence (measured as $\alpha_1+\alpha_2+\beta+\gamma/2 +\delta_2/13$) estimated around $0.96$ across the considered assets. As expected, a large part of volatility comes from the ARCH terms, with $\hat\alpha_1$ ranging between $0.22$ and $0.27$ and $\hat\alpha_2$ between $-0.11$ and $-0.21$, given that, at an intradaily level, volatility is very sensitive to news (with the first lag), but then contributes (with the second lag) to the absorption of news \citep[see, for example][]{Cipollini:Gallo:Palandri:2020}. This is also confirmed by the coefficients $\gamma$, which measure the impact of bad news (as represented by negative returns): they are positive and significant at a 1\% level, with values between $0.01$ and $0.03$. The positiveness of $\delta_1$ represents evidence in favour of the  \textit{market opening effect}, i.e. a response of volatility at the beginning of each trading day between $0.008$ and $0.017$, due to whatever accumulation of news entails an overnight price movement. By the same token, the other bin--specific effect, namely the effect of the first bin on the second bin (marked by $\delta_2$), enters the model with the expected (and significant) negative sign, that is, the contribution to current bin volatility from the previous one is lower for bins after the first one. In such a case, the total effect is given by $\alpha_1+\delta_2$, ranging between 0.18 and 0.23 across the considered series. This is in line with what is shown in Figure \ref{fig:intraday_mean}: since volatility is systematically higher in the first bin, the dynamics in the subsequent bins pays less attention, so to speak, to the previous bin value.

As for the jump component, the autoregressive coefficient $\varphi$ is significant hovering around $0.3$ across tickers, pointing out to a low persistence of the jump component - in line with the common view that volatility jumps are short-lived occurrences. Given our main hypothesis about the relationship between observed and expected jumps, the coefficient $\psi$ is positive, with expected jumps that are an increasing function of the identified significant jumps.  

\begin{figure}[t]
	\begin{center}
		\subfigure[MSFT]{\includegraphics[height=6.5cm,width=8cm]{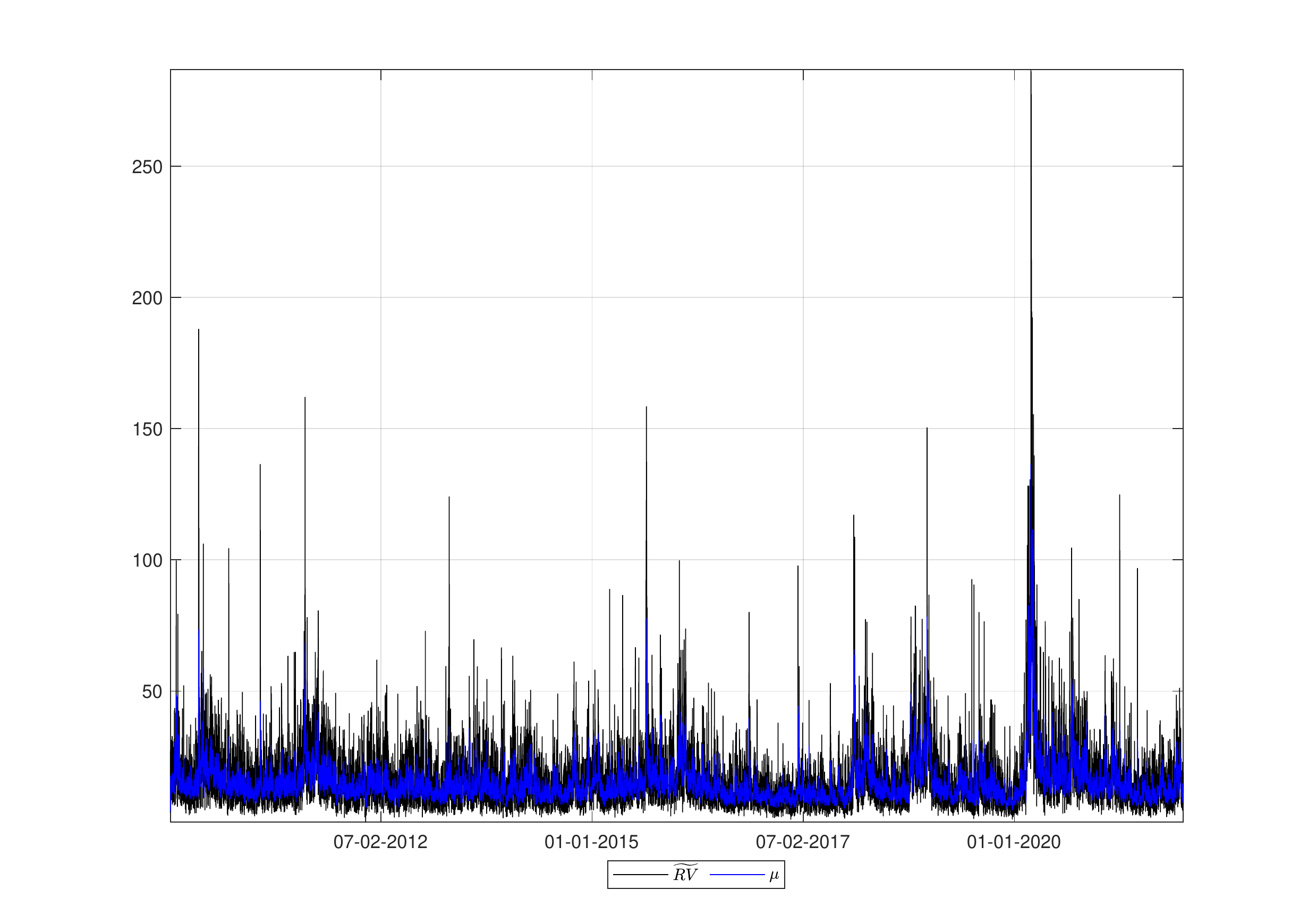}}
		\subfigure[JPM]{\includegraphics[height=6.5cm,width=8cm]{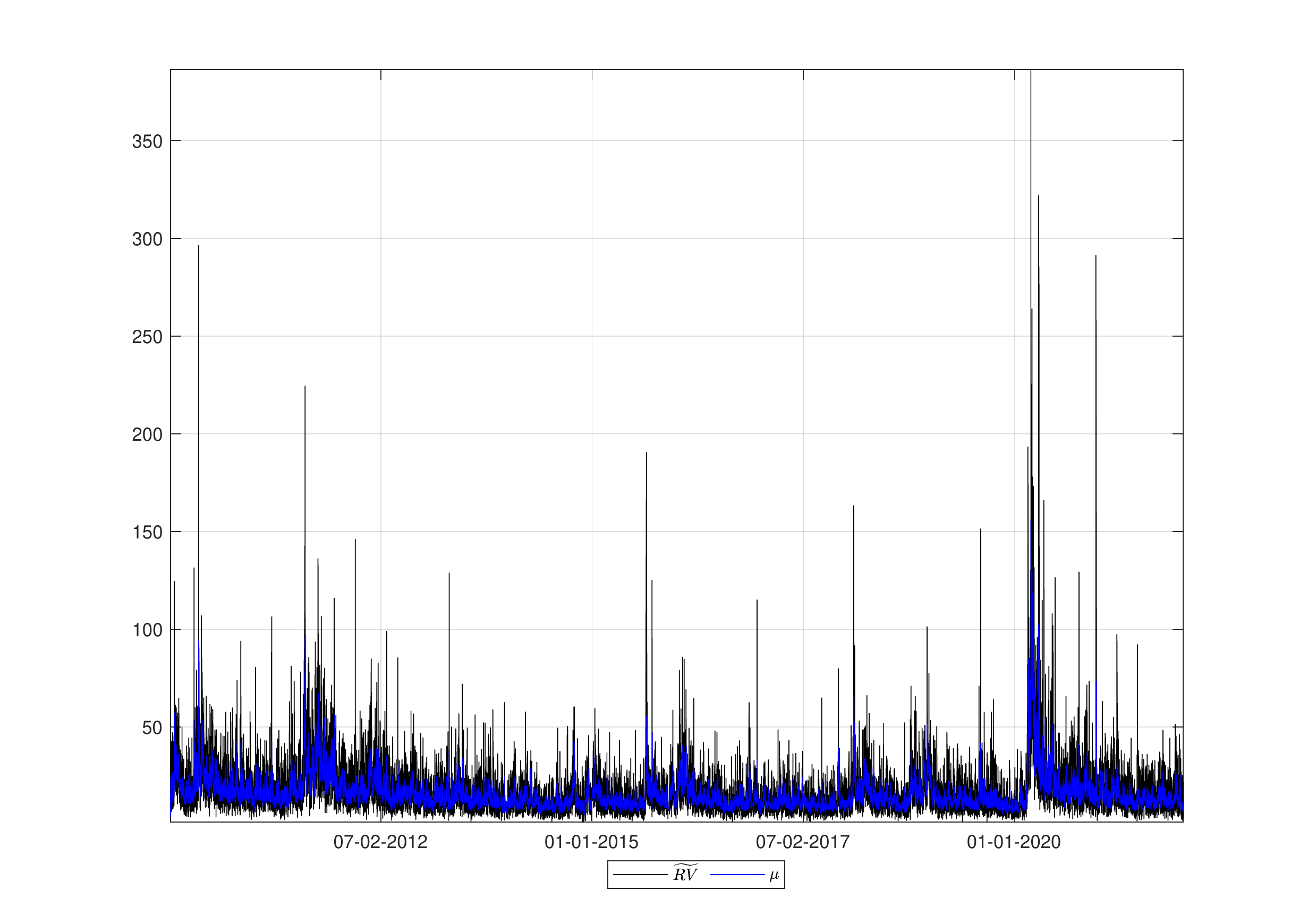}}\\
		\subfigure[CAT]{\includegraphics[height=6.5cm,width=8cm]{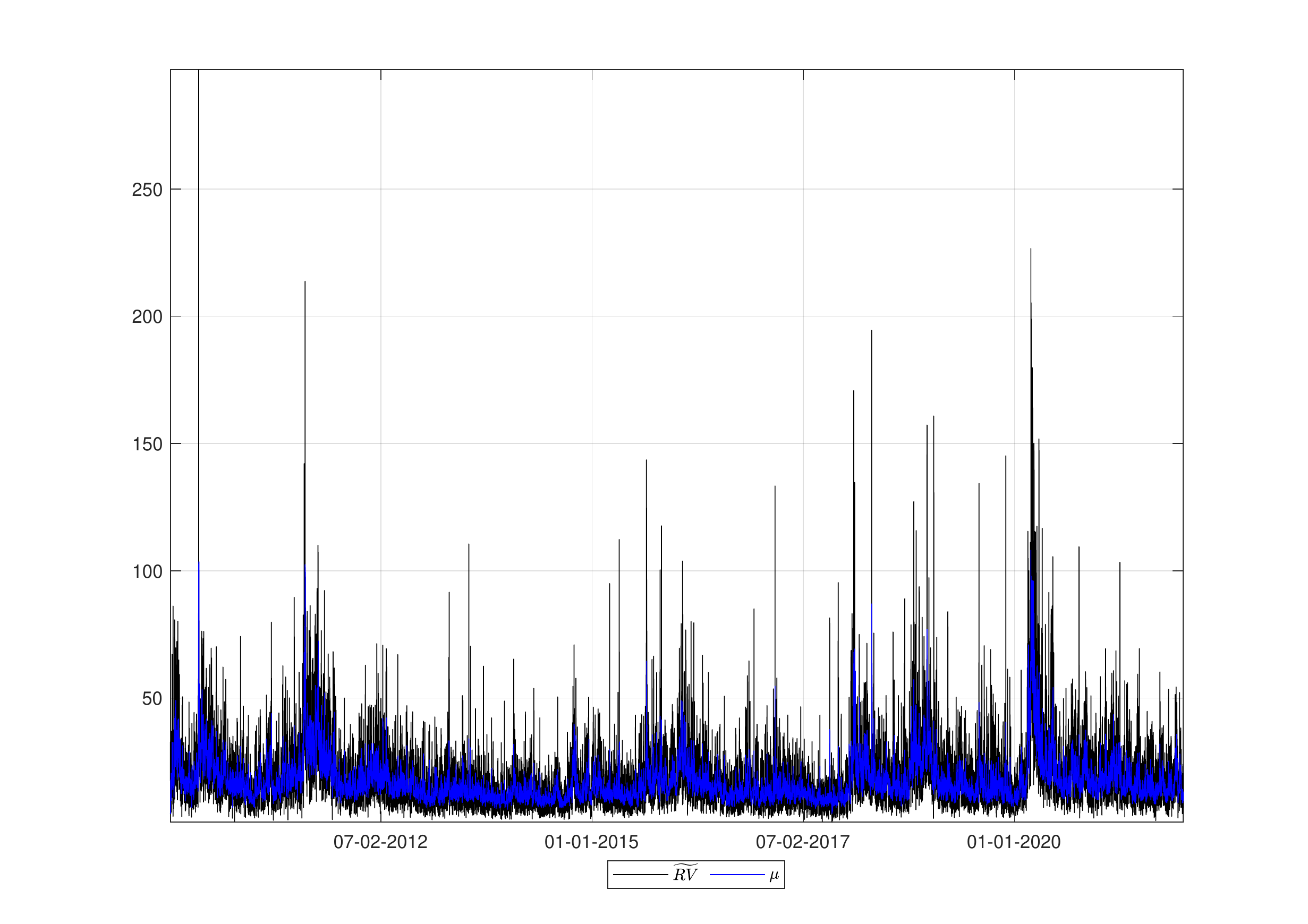}}		\subfigure[HD]{\includegraphics[height=6.5cm,width=8cm]{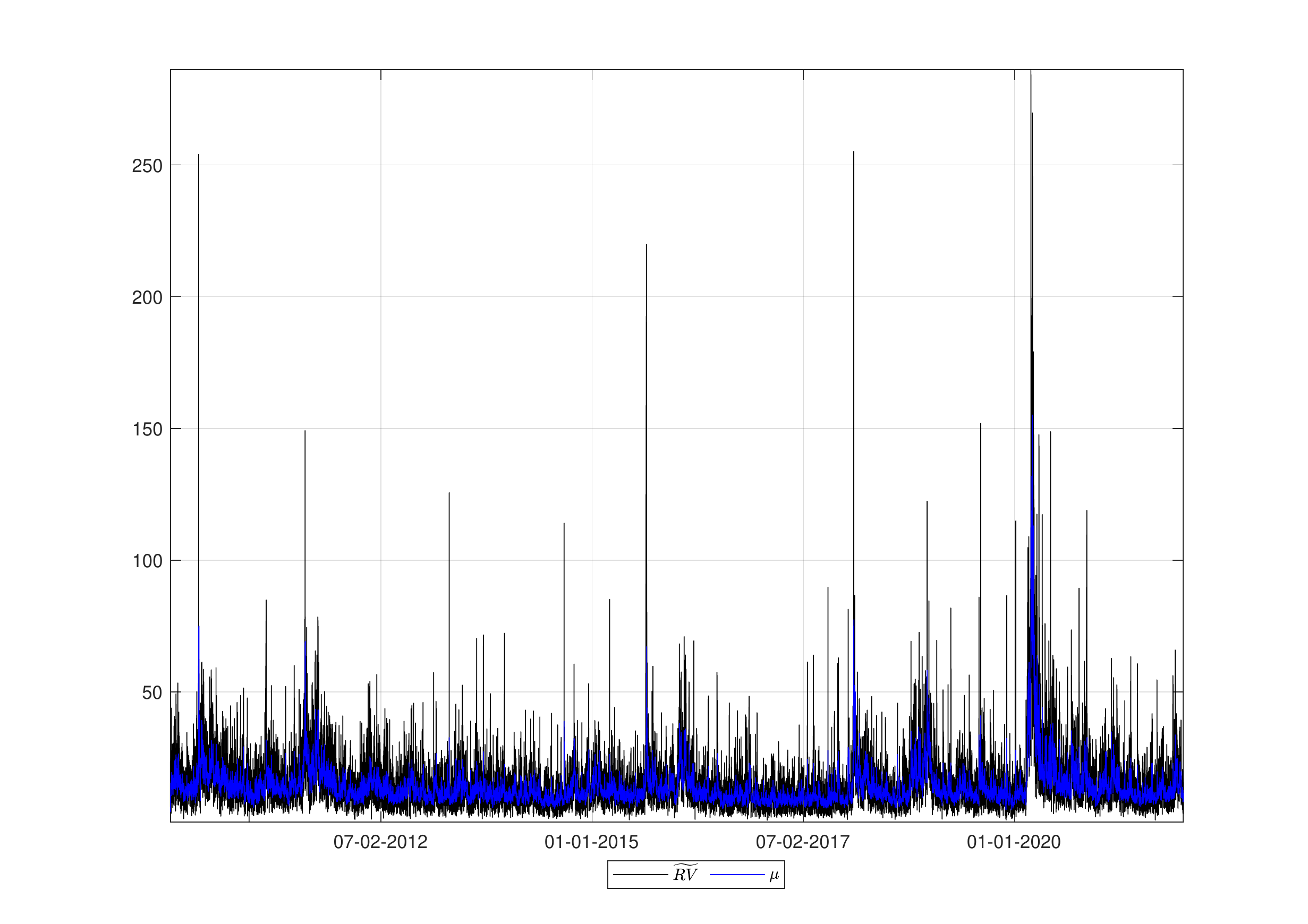}}
		\caption{Deseasonalized realized volatility ($\widetilde{RV}_{i,t}$, black line) and conditional expectation ($\mu_{i,t}$, blue line) for MSFT, HD, JPM and CAT. Model: restricted AJM. Sample period: January 3, 2010 -- December 31, 2021.\label{fig:rv_mu}}
	\end{center}
\end{figure}

Figure \ref{fig:rv_mu} shows the evolution of $\widetilde{RV}_{i,t}$ (black line) together with its conditional expectation $\mu_{i,t}$ (blue line) as estimated from the restricted model for MSFT, HD, JPM and CAT. In all the cases, the series follow a similar path with $\mu_{i,t}$ being able of reproducing the spikes characterizing the observed volatility series. Moreover, even the conditional mean is characterized by volatility clustering, with high volatility corresponding to turbulent periods of the market (see discussion is Section \ref{sec:dataset}).

Panel b) of Table \ref{tab:estimation_results} shows the p--value of the Ljung-Box test statistics on residuals. The richer model manages to capture some of the residual autocorrelation which is present in the restricted model (where $\alpha_2=0$). Even with this model (in the first column), estimation results remain unchanged in terms of sign and significance of the estimated parameters, with a slight worsening of the fitting properties.  For what matters later, the inclusion or exclusion of the second lag of $C_{i,t}$ does not affect the classification results, suggesting that residual autocorrelation does not have an impact for clustering purposes. 
	
	\section{The classification method}
	\label{sec:classification}
	
	Classifying monetary policy announcements is important for both investors and policy makers who can employ useful information to predict the reaction of volatility to a Central Bank's communication. Our model suggests a simple mechanism for a model-based classification of monetary announcements according to their impact on volatility jumps, that is, we distinguish expected jumps (as represented by $\kappa_{i,t}$ in Eq. \ref{mod:amem_jump}) from what we call \textit{jump surprise} or \textit{unexpected} jumps. The latter is defined as the difference between what we expect and what we observe, $$J^{surprise}_{\iota,\tau}=\kappa_{\iota,\tau}-SJ_{\iota,\tau}$$ with $\tau=t$ on announcements days and $\iota$ indicating the first bin past the announcement time. 

In line with the already mentioned time-point classification methods, focusing on the $\tau$ announcement days, we suggest to classify the $104$ Fed's announcements, according to whether at the bin $\iota$ within that day we detect 
\begin{enumerate}
	\item an Upward Spike  (local maximum), if $\kappa_{\iota-1,\tau}<\kappa_{\iota,\tau}>\kappa_{\iota+1,\tau}$;
	\item a  Downward Spike (local minimum), if $\kappa_{\iota-1,\tau}>\kappa_{\iota,\tau}<\kappa_{\iota+1,\tau}$;
	\item a  Boost (an increase), if $\kappa_{\iota-1,\tau}<\kappa_{\iota,\tau}$;
	\item a  Drop (a reduction), if $\kappa_{\iota-1,\tau}>\kappa_{\iota,\tau}$,
\end{enumerate}
for $\kappa$ (and, similarly, for $J^{surprise}$). Note that the identification of the Boost or the Drop bins of the announcement day is made excluding the corresponding spikes.

Our classification method has the merit of being immediately applicable when an announcement is released and has the unique characteristic of giving information about the expected jumps. 
The impact of an announcement should affect the market in similar ways across stocks or perhaps affect some sectors more than others: to that end, we evaluate the agreement about the classification obtained ticker by ticker, resorting to the adjusted Rand-index \citep{Rand:1971,Hubert:Arabie:1985} computed between pairs of assets. This index ranges in the interval [0,1], taking value of 0 in the case of maximum difference between two methods, while it is equal to 1 in the case of perfect matching. 

In addition, in order to exploit all the information content of the expected jump series, we propose to evaluate our method relying also on the similarity between the classification based on $\kappa_{\iota,\tau}$ with respect to the classification based on $J^{surprise}_{\iota,\tau}$. By so doing, it is possible to assess the ability of the model to anticipate a jump surprise. This could represent a matter of interest for investors who design trading strategy based on volatility (e.g. short and long positions based on volatility options). Moreover, correctly predicting volatility to be in either the \textit{Upward Spike} or \textit{Downward Spike} cluster might be useful in constructing the well-known \textit{momentum} strategy.
	
	\subsection{Monetary announcements classification}
	\label{sec:classification_results}
	
	The classification results are shown in Table \ref{tab:classification}. The first panel is devoted to the expected jumps $\kappa_{\iota,\tau}$: most of the announcements (around 37\%) correspond to a Downward Spike, followed by Upward Spikes (between 20\% and 26\% of the cases), and  Boosts (between 12\% and 20\%). Results are quite homogeneous across assets, even if only few announcements belong to the same cluster for all the considered tickers. This is the case of the announcements -- classified as Downward Spike -- released on October 30, 2013 and January 30, 2019, while the announcement on April 30, 2014 (communicating an increase in the assets purchases program) is related to a reduction (Drop) of $\kappa$. 

\begin{table}[t]
	\footnotesize
	\caption{Classification results based on the restricted AJM. 104 Fed's monetary announcements are classified basing on their effects either on $\kappa_{\iota,\tau}$ or on the jump surprise, $J^{Surprise}_{i,t}$.\label{tab:classification}}
	\begin{adjustbox}{max width=1\linewidth,center}
		\begin{tabular}{lcccc|cccc}
			\multicolumn{9}{c}{$\kappa_{\iota,\tau}$}\\
			&\multicolumn{4}{c}{Monetary Announcements}  &\multicolumn{4}{c}{Forward Guidance}       \\[2mm]
			& Upward Spike & Downward Spike & Boost & Drop & Upward Spike & Downward Spike & Boost & Drop\\[2mm]
			\midrule
			MSFT & 30 & 27 & 27 & 19 & 12 & 6  & 16 & 13 \\
			GS   & 30 & 27 & 30 & 17 & 9  & 14 & 20 & 4  \\
			JPM  & 28 & 35 & 28 & 13 & 16 & 17 & 11 & 3  \\
			JNJ  & 33 & 30 & 30 & 11 & 17 & 11 & 16 & 3  \\
			CAT  & 29 & 39 & 26 & 10 & 15 & 17 & 11 & 4  \\
			MMM  & 35 & 35 & 21 & 13 & 18 & 15 & 10 & 4  \\
			HD   & 25 & 39 & 26 & 14 & 12 & 13 & 13 & 9 \\
			\midrule
			\multicolumn{9}{c}{$J^{surprise}_{\iota,\tau}$}        \\[2mm]
			&\multicolumn{4}{c}{Monetary Announcements}  &\multicolumn{4}{c}{Forward Guidance}       \\[2mm]
			& Upward Spike & Downward Spike & Boost & Drop & Upward Spike & Downward Spike & Boost & Drop\\
			MSFT & 32 & 45 & 5  & 21 & 14 & 17 & 4 & 12 \\
			GS   & 34 & 39 & 11 & 20 & 10 & 24 & 7 & 6  \\
			JPM  & 29 & 46 & 9  & 20 & 16 & 20 & 5 & 6  \\
			JNJ  & 34 & 51 & 5  & 14 & 16 & 25 & 3 & 3  \\
			CAT  & 32 & 47 & 10 & 15 & 15 & 22 & 5 & 5  \\
			MMM  & 34 & 45 & 4  & 21 & 17 & 18 & 3 & 9  \\
			HD   & 32 & 43 & 5  & 24 & 16 & 16 & 3 & 12\\
			\bottomrule
		\end{tabular}
	\end{adjustbox}
\end{table}

Results remain almost unchanged when dealing with jump surprise, with 2 announcements (concerning decisions either on FFR or assets purchases program) having a common classification for all the assets. 

\begin{figure}[t]
	\subfigure[MSFT]{\includegraphics[height=2.8cm,width=8cm]{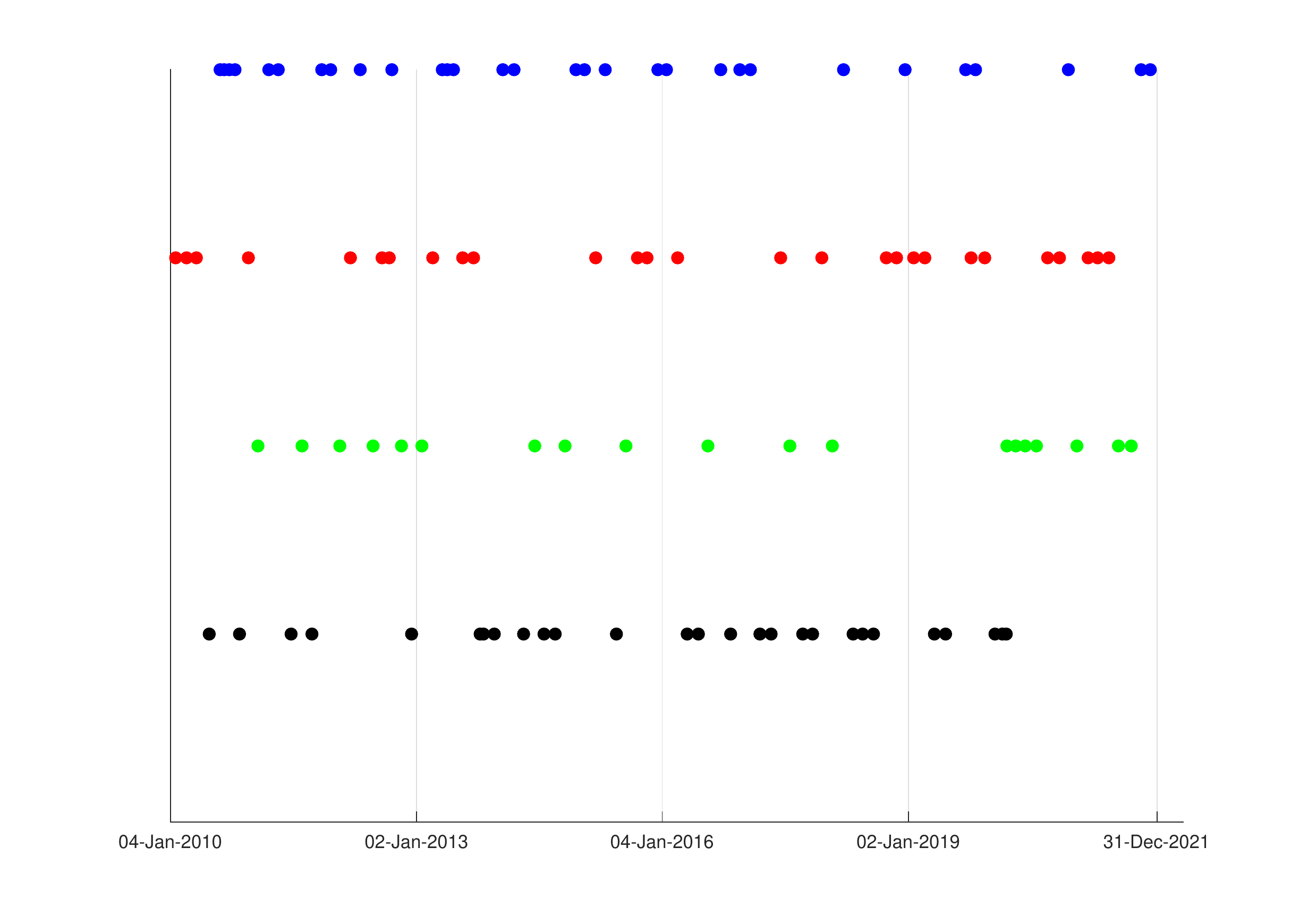}}
	\subfigure[GS]{\includegraphics[height=2.8cm,width=8cm]{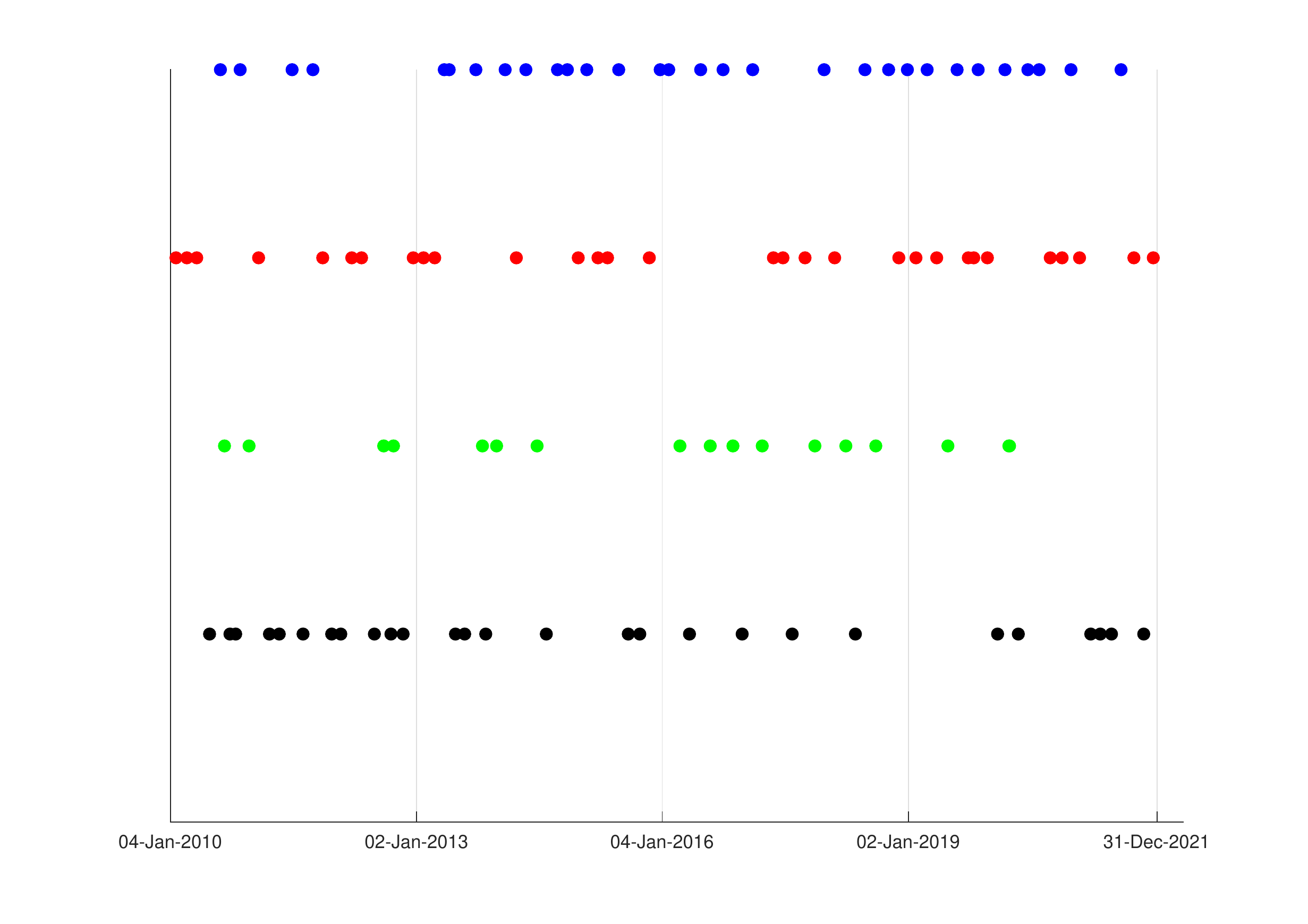}}\\
	\subfigure[JPM]{\includegraphics[height=2.8cm,width=8cm]{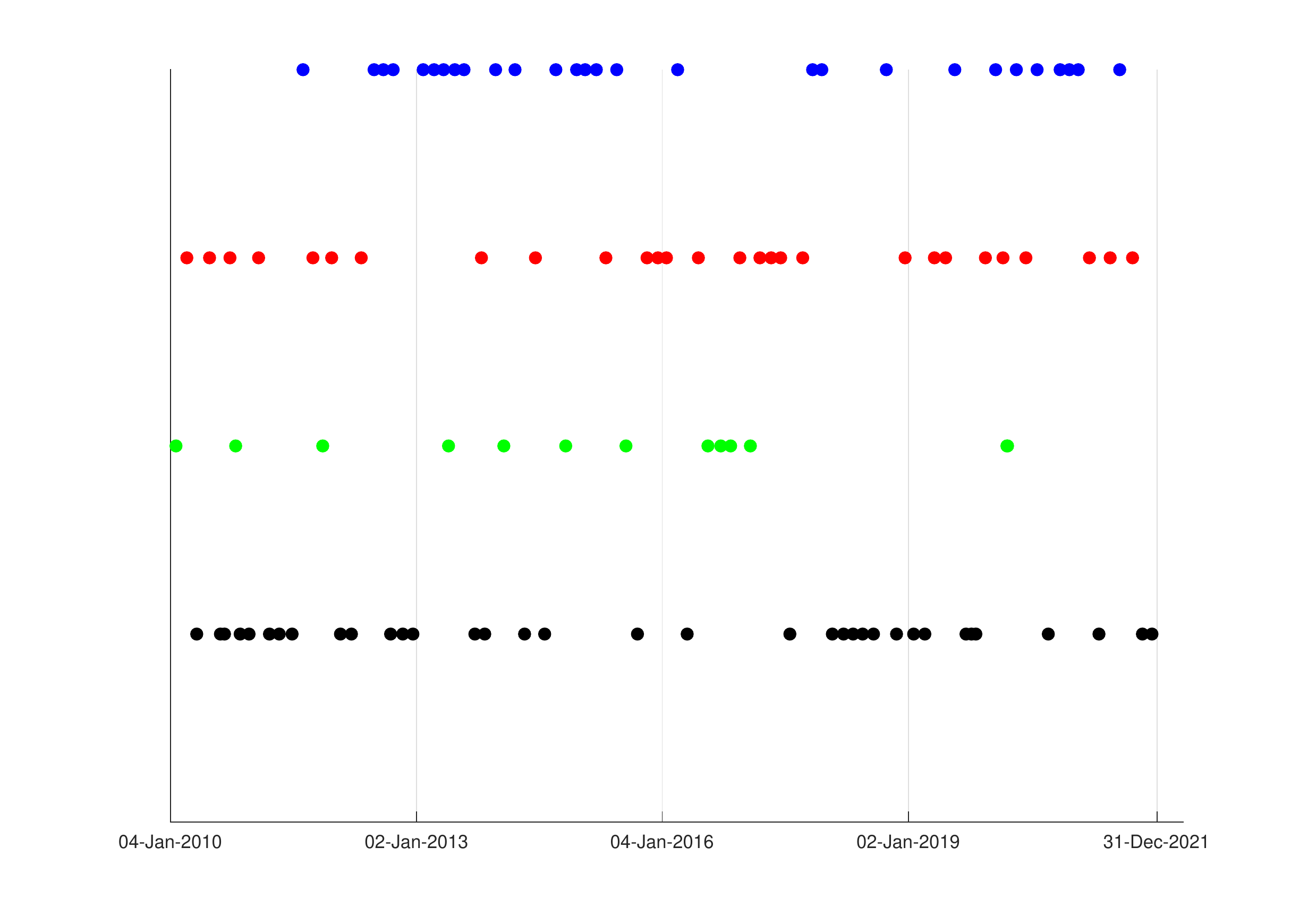}}
	\subfigure[JNJ]{\includegraphics[height=2.8cm,width=8cm]{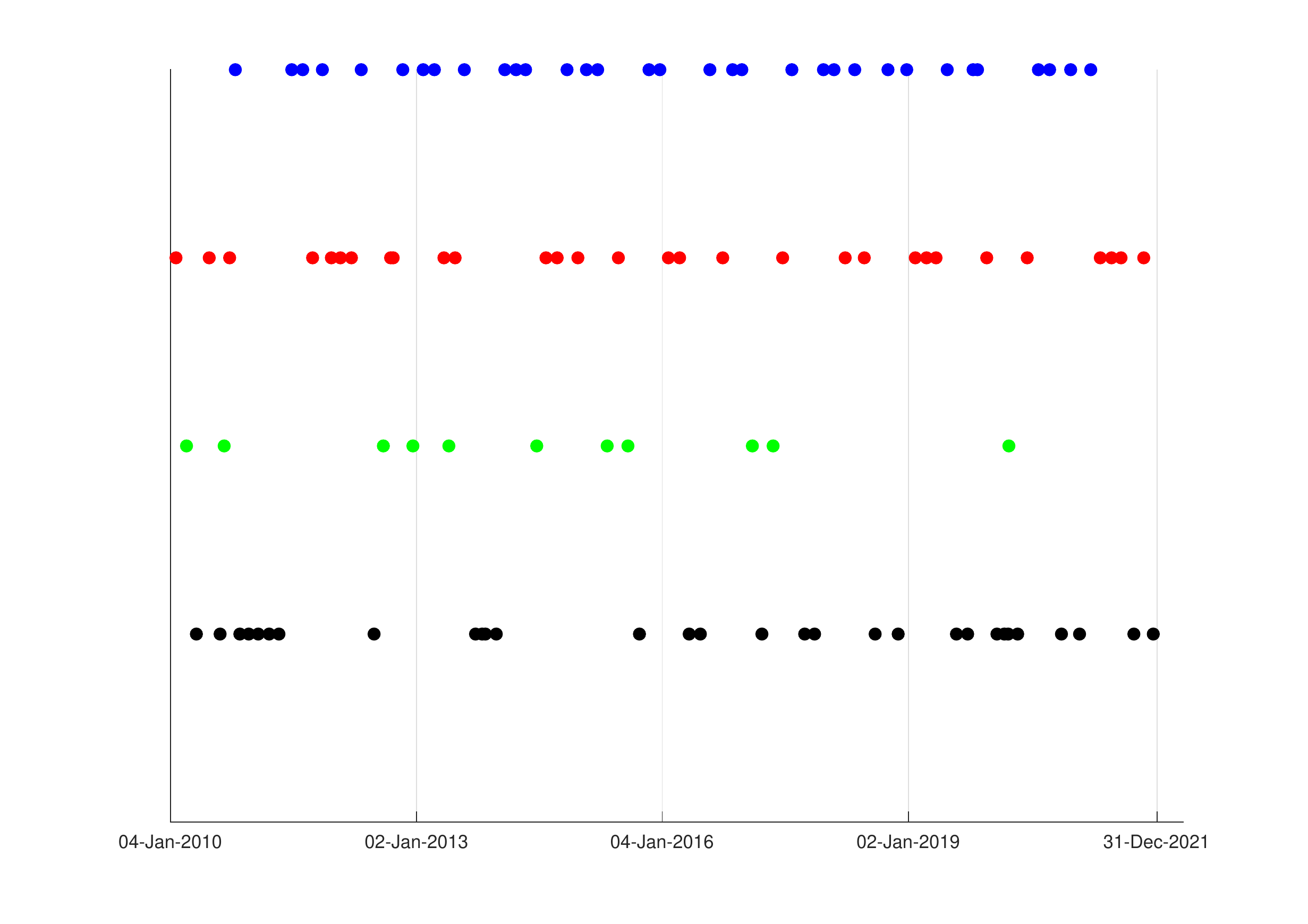}}\\
	\subfigure[CAT]{\includegraphics[height=2.8cm,width=8cm]{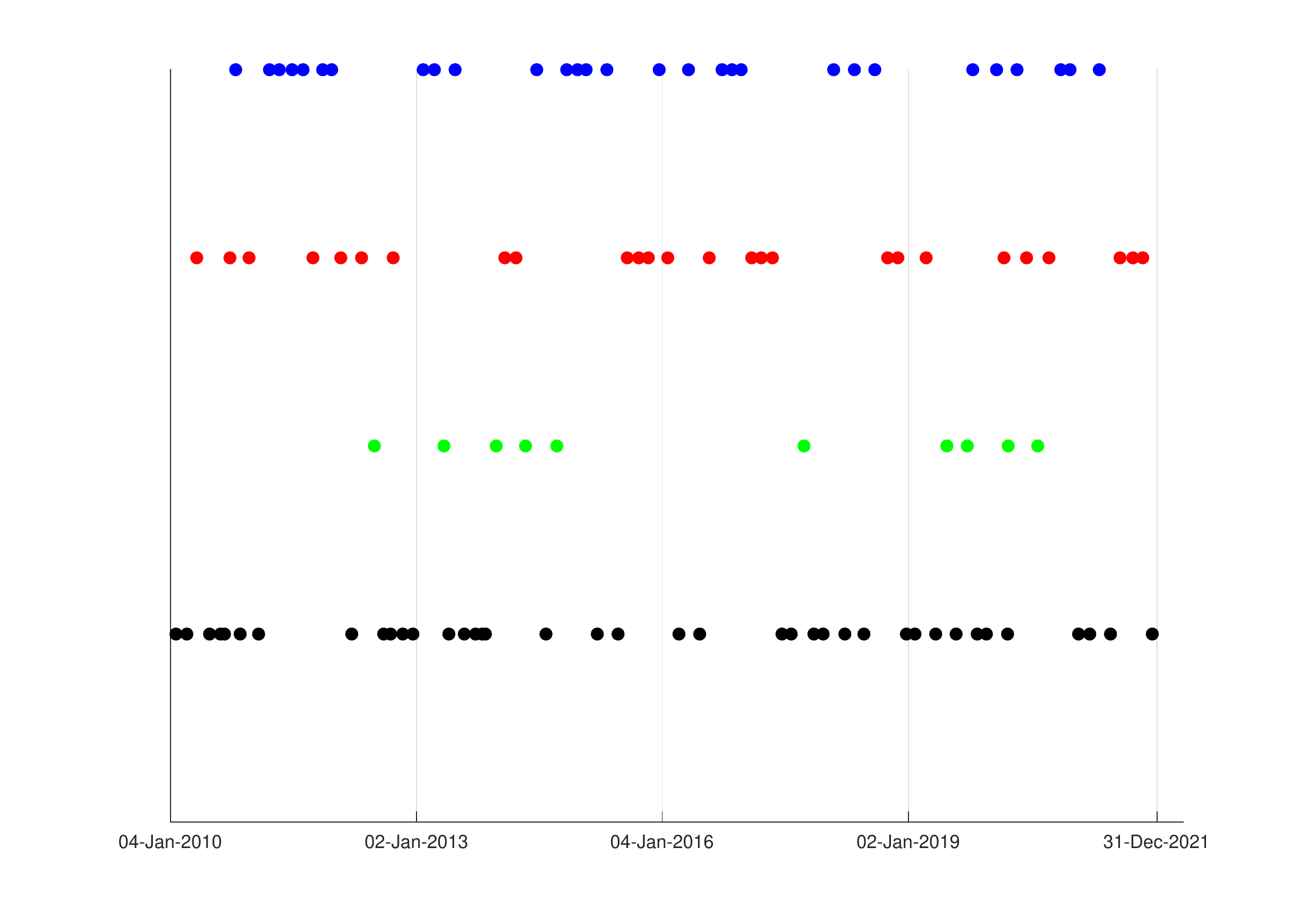}}
	\subfigure[MMM]{\includegraphics[height=2.8cm,width=8cm]{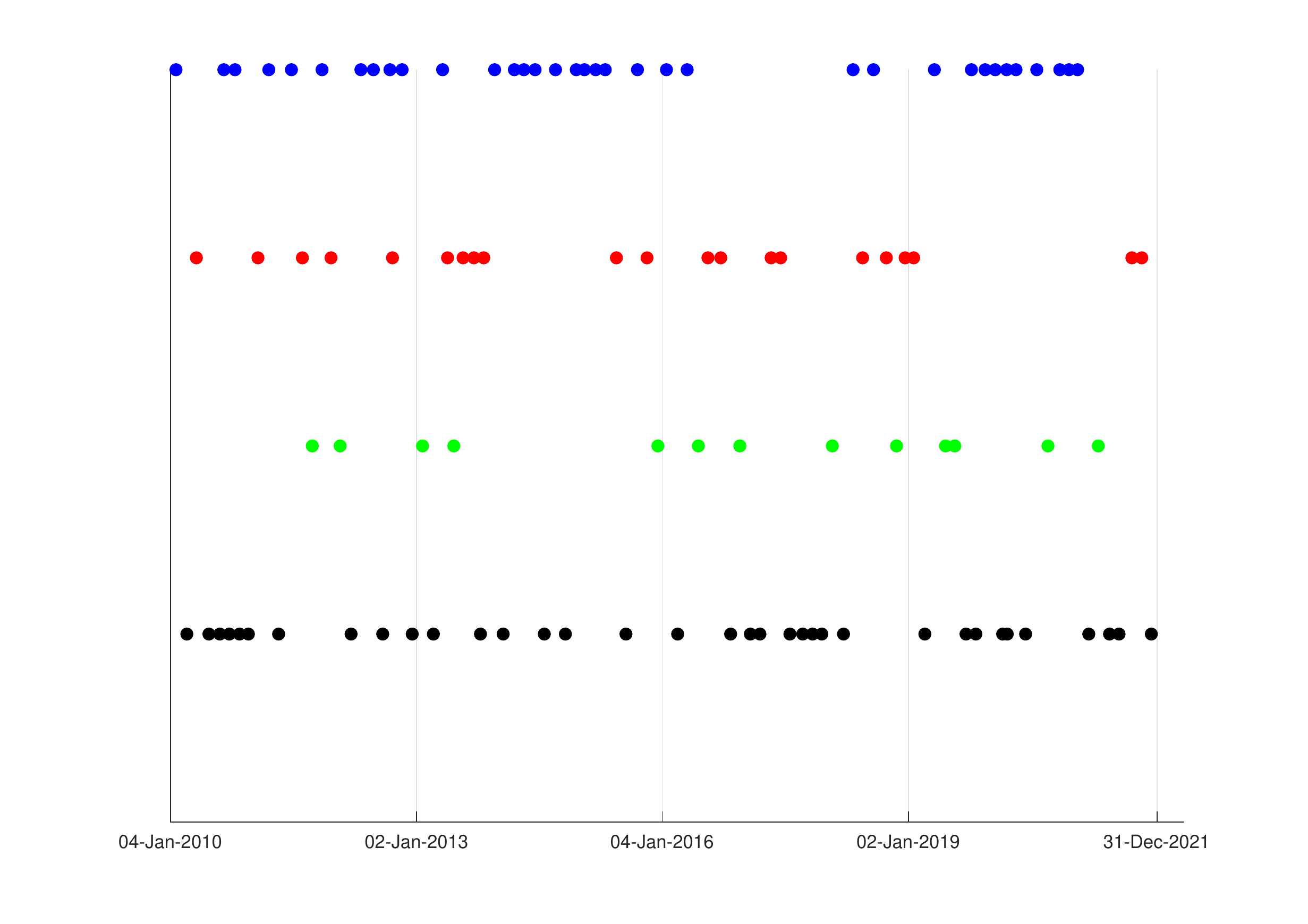}}\\
	\subfigure[HD]{\includegraphics[height=2.8cm,width=8cm]{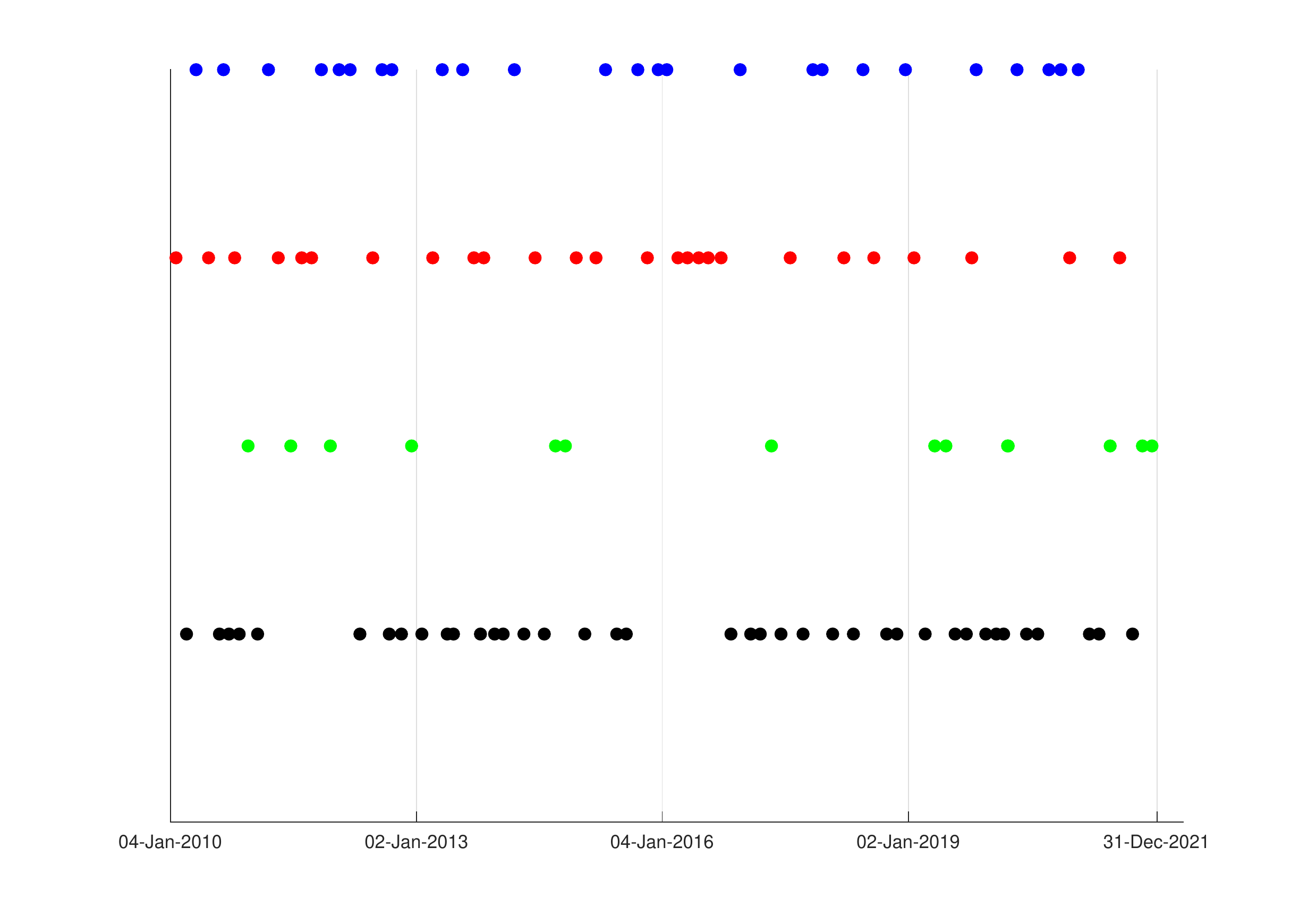}}
	\caption{Monetary announcements classification based on expected jump $\kappa$: Upward Spike (solid blue dots), Downward Spike (black dots), Boost (red dots) and Drop (green dots). Sample period: January 3, 2010 -- December 31, 2021.\label{fig:clustering}}
\end{figure}

Interestingly, some announcements affected only specific sectors. Considering the results about expected jumps (refer to Figure \ref{fig:clustering} for a visual evidence of the comments that follow), for example, the decisions about the FFR in certain dates (June 17, 2015; July 31, 2019; July 28, 2021) are classified as Upward Spike only for the financial sector. Conversely, some Fed communications had a different impact in different sectors, with announcements on April 27, 2016, August 1, 2018 causing a Upward Spike for the industrial sectors (as represented by CAT and MMM) while they had the opposite effect (Downward Spike) on the financial and Information Technology sectors. As expected, both the pharmaceutical (JNJ) and the customer discretionary (HD) sectors are not integrated with the others, with around the 12\%, respectively, 20\% of announcements that have a different classification with respect to the other securities. For the customer discretionary sector, in particular, the 42\% of announcements classified as Drop has a different classification when another stock is considered. This does not hold for the IT sector (MSFT), which shares the 27\% of the classification with the financial sector -- the percentage is 26\% (31\%) for announcements classified as Upward Spike (Downward, respectively). For an investor, the usefulness of this kind of information is twofold: on the one hand, knowing the degree of interconnection between sectors is important in designing diversification strategy; on the one other, predicting turning points of $\kappa_{\iota,\tau}$ is crucial for investment strategies based on momentum. 
When one turns to jump surprises, a higher degree of integration is encountered for the pharmaceutical and customer discretionary sector with 37\% classified as Drop only for JNJ and HD. Similarly, the integration between the IT and financial sectors increases, with 35\% of announcements having a common classification for MSFT and JPM -- the percentage is 33\% (62\%) if the Upward Spike (Downward, respectively) is considered.

\begin{table}[th]
	\centering
	\caption{Panel \textbf{a)} reports the percentage adjusted Rand-index between pairs of assets basing either on $\kappa$ (lower triangular matrix) or $J^{Surprise}$ (upper triangular matrix); panel \textbf{b)} refers to the percentage adjusted Rand-index between $\kappa$ and  $J^{Surprise}$; panel \textbf{c)} shows the percentage adjusted Rand-index between $\kappa$ and $SJ$.  \label{tab:rand_index}}
	\begin{adjustbox}{max width=1\linewidth,center}
		\begin{tabular}{lccccccc}
			\toprule
			\textbf{a)}		& MSFT    & GS      & JPM     & JNJ     & CAT     & MMM     & HD      \\
			MSFT          & -     & 57.67\% & 56.70\% & 55.06\% & 57.37\% & 55.73\% & 57.04\% \\
			GS            & 63.13\% & -     & 58.78\% & 56.46\% & 56.91\% & 56.68\% & 57.39\% \\
			JPM           & 62.60\% & 61.54\% & -     & 55.04\% & 56.39\% & 57.13\% & 57.66\% \\
			JNJ           & 61.67\% & 61.31\% & 61.09\% & -     & 55.97\% & 56.42\% & 55.53\% \\
			CAT           & 62.79\% & 60.53\% & 60.38\% & 60.34\% & -     & 60.72\% & 57.43\% \\
			MMM           & 61.24\% & 60.51\% & 60.36\% & 59.65\% & 62.08\% & -     & 57.80\% \\
			HD            & 62.60\% & 60.87\% & 60.46\% & 59.93\% & 59.67\% & 60.29\% & -     \\
			\hline 
			\textbf{b)} &  81.11\% & 75.58\% & 73.94\% & 77.22\% & 79.28\% & 79.37\% & 74.42\%  \\
			\textbf{c)}& 63.24\% & 56.55\% & 59.48\% & 62.47\% & 59.39\% & 61.82\% & 57.99\% \\
			\bottomrule
		\end{tabular}
	\end{adjustbox}
\end{table}
\pagebreak
To be more rigorous in evaluating the accuracy of our classification method, we rely on the adjusted Rand-index computed between pairs of assets. Table \ref{tab:rand_index} shows results deriving either on $\kappa$ or $J^{Surprise}$ (arranged pairwise in the upper, respectively lower, triangular matrix). Consistently, elements of the lower portion (expected jumps) are pairwise larger than the corresponding elements of the upper portion (surprise jumps), with indices that are close to 60\%: since these values are away from zero a similar clustering among tickers is apparent, but an idiosyncratic component of reaction to announcements is present. This is confirmed by the fact that the index is higher than 75\% when $\kappa$ is compared to $J^{surprise}$, while it 63\% at most, when we compare $\kappa$ to SJ. 

	\section{Concluding remarks}
	\label{sec:conclusion}
	A novel Multiplicative Error Model was suggested here to decouple the continuous part of volatility from its jump component, geared toward a model--based classification of Central Bank's announcements based on their impact on volatility jumps. This approach is different from previous analyses aimed at evaluating the impact of monetary announcements on volatility: first and foremost, we reconstruct the dynamics of intradaily volatility by thirty-minute bins distinguishing a continuous part and a jump part components and accommodating some peculiarities related to the time--of--day effect. We allow for a time-varying announcement effect, with each Central Bank's communication having its own effect on volatility. This allows us to reconstruct expected jumps and jump surprises, and then classify monetary announcements according to whether we detect a local maximum (minimum) or a simple increment (reduction) of the considered series on announcement days. 

We apply our approach to seven US tickers focusing on the Fed's monetary policy announcements and we highlight the importance of its policy decisions for the evolution of stock volatility and the dynamics of expected jumps. The results on announcement days classification reveal both commonalities and differences among different sectors of the market, with a high interconnection between the IT and financial sectors, with the pharmaceutical and the customer discretionary sectors being less close. This has the potential to provide useful information to both policy makers -- who might acquire new knowledge about the financial implications of their policies -- and investors -- for their investment and risk management decisions.

The evaluation of our method, based either on the adjusted Rand-index or on the correspondence between expected and observed jumps, reveals a good accuracy of our classification, which has the merit of being immediately applicable when an announcement is released. 

The model could be further refined by extending the number of lags in the specification to check whether it improves the autocorrelation properties of the residuals or it has an impact on the subsequent classification; moreover, a feedback effect could be accommodated within the dynamics of the continuous part making it depend on the lagged expected jump component. In the specification, in line with the logic of the Composite MEM, the two components enter the model additively, while an interesting comparison would be with a model with multiplicative components.

	\bibliographystyle{chicago}
	\bibliography{biblio.bib}

\begin{thebibliography}{}

\bibitem[\protect\citeauthoryear{Aghabozorgi, Shirkhorshidi, and
  Wah}{Aghabozorgi et~al.}{2015}]{Aghabozorgi:Shirkhorshidi:Wah:2015}
Aghabozorgi, S., A.~S. Shirkhorshidi, and T.~Y. Wah (2015).
\newblock Time-series clustering--a decade review.
\newblock {\em Information systems\/}~{\em 53}, 16--38.

\bibitem[\protect\citeauthoryear{Andersen, Bollerslev, and Diebold}{Andersen
  et~al.}{2007}]{Andersen:Bollerslev:Diebold:2007}
Andersen, T.~G., T.~Bollerslev, and F.~X. Diebold (2007).
\newblock Roughing it up: Including jump components in the measurement,
  modeling and forecasting of return volatility.
\newblock {\em Review of Economics and Statistics\/}~{\em 89}, 701--720.

\bibitem[\protect\citeauthoryear{Andersen, Bollerslev, and Diebold}{Andersen
  et~al.}{2010}]{Andersen:Bollerslev:Diebold:2010}
Andersen, T.~G., T.~Bollerslev, and F.~X. Diebold (2010).
\newblock Parametric and nonparametric volatility measurement.
\newblock In {\em Handbook of financial econometrics: Tools and techniques},
  pp.\  67--137. Elsevier.

\bibitem[\protect\citeauthoryear{Andersen, Bollerslev, Diebold, and
  Vega}{Andersen et~al.}{2007}]{Andersen:Bollerslev:Diebold:Vega:2007}
Andersen, T.~G., T.~Bollerslev, F.~X. Diebold, and C.~Vega (2007).
\newblock Real-time price discovery in global stock, bond and foreign exchange
  markets.
\newblock {\em Journal of International Economics\/}~{\em 73\/}(2), 251--277.

\bibitem[\protect\citeauthoryear{Anderson, Bollerslev, Diebold, and
  Vega}{Anderson et~al.}{2003}]{Andersen:Bollerslev:Diebold:Vega:2003}
Anderson, T.~G., T.~Bollerslev, F.~X. Diebold, and C.~Vega (2003).
\newblock Micro effects of macro announcements: Real-time price discovery in
  foreign exchange.
\newblock {\em American Economic Review\/}~{\em 93\/}(1), 38--62.

\bibitem[\protect\citeauthoryear{Barndorff-Nielsen and
  Shephard}{Barndorff-Nielsen and
  Shephard}{2004a}]{BarndorffNielsen:Shephard:2004}
Barndorff-Nielsen, O. and N.~Shephard (2004a).
\newblock Power and bipower variation with stochastic volatility and jumps
  (with discussion).
\newblock {\em Journal of Financial Econometrics\/}~{\em 2}, 1--48.

\bibitem[\protect\citeauthoryear{Barndorff-Nielsen and
  Shephard}{Barndorff-Nielsen and
  Shephard}{2004b}]{BarndorffNielsen:Shepard:2004b}
Barndorff-Nielsen, O.~E. and N.~Shephard (2004b).
\newblock How accurate is the asymptotic approximation to the distribution of
  realized variance.
\newblock {\em Identification and inference for econometric models. A
  Festschrift in honour of TJ Rothenberg\/}, 306--311.

\bibitem[\protect\citeauthoryear{Barndorff-Nielsen and
  Shephard}{Barndorff-Nielsen and
  Shephard}{2006}]{BarndorffNielsen:Shephard:2006}
Barndorff-Nielsen, O.~E. and N.~Shephard (2006).
\newblock Econometrics of testing for jumps in financial economics using
  bipower variation.
\newblock {\em Journal of Financial Econometrics\/}~{\em 4}, 1--30.

\bibitem[\protect\citeauthoryear{Bollerslev}{Bollerslev}{1986}]{Bollerslev:1986}
Bollerslev, T. (1986).
\newblock Generalized autoregressive conditional heteroskedasticity.
\newblock {\em Journal of Econometrics\/}~{\em 31}, 307--327.

\bibitem[\protect\citeauthoryear{Bomfim}{Bomfim}{2003}]{Bomfim:2003}
Bomfim, A.~N. (2003).
\newblock Pre-announcement effects, news effects, and volatility: monetary
  policy and the stock market.
\newblock {\em Journal of Banking \& Finance\/}~{\em 27\/}(1), 133--151.

\bibitem[\protect\citeauthoryear{Brownlees, Cipollini, and Gallo}{Brownlees
  et~al.}{2012}]{Brownlees:Cipollini:Gallo:2012}
Brownlees, C.~T., F.~Cipollini, and G.~M. Gallo (2012).
\newblock Multiplicative error models.
\newblock In L.~Bauwens, C.~Hafner, and S.~Laurent (Eds.), {\em Volatility
  Models and Their Applications}, pp.\  223--247. Wiley.

\bibitem[\protect\citeauthoryear{Brownlees and Gallo}{Brownlees and
  Gallo}{2006}]{Brownlees:Gallo:2006}
Brownlees, C.~T. and G.~M. Gallo (2006).
\newblock Financial econometric analysis at ultra{--}high frequency: Data
  handling concerns.
\newblock {\em Computational Statistics and Data Analysis\/}~{\em 51},
  2232--2245.

\bibitem[\protect\citeauthoryear{Busch, Christensen, and Nielsen}{Busch
  et~al.}{2011}]{Busch:Christensen:Nielsen:2011}
Busch, T., B.~J. Christensen, and M.~{\O}. Nielsen (2011).
\newblock The role of implied volatility in forecasting future realized
  volatility and jumps in foreign exchange, stock, and bond markets.
\newblock {\em Journal of Econometrics\/}~{\em 160\/}(1), 48--57.

\bibitem[\protect\citeauthoryear{Caiado and Crato}{Caiado and
  Crato}{2007}]{Caiado:Crato:2007}
Caiado, J. and N.~Crato (2007).
\newblock A {GARCH}-based method for clustering of financial time series:
  International stock markets evidence.
\newblock In {\em Recent advances in stochastic modeling and data analysis},
  pp.\  542--551. World Scientific.

\bibitem[\protect\citeauthoryear{Caporin, Rossi, and Santucci
  De~Magistris}{Caporin et~al.}{2017}]{Caporin:Rossi:Santucci:2017}
Caporin, M., E.~Rossi, and P.~Santucci De~Magistris (2017).
\newblock Chasing volatility: a persistent multiplicative error model with
  jumps.
\newblock {\em Journal of Econometrics\/}~{\em 198}, 122--145.

\bibitem[\protect\citeauthoryear{Cipollini, Engle, and Gallo}{Cipollini
  et~al.}{2013}]{Cipollini:Engle:Gallo:2013}
Cipollini, F., R.~F. Engle, and G.~M. Gallo (2013).
\newblock Semiparametric vector {MEM}.
\newblock {\em Journal of Applied Econometrics\/}~{\em 28}, 1067--1086.

\bibitem[\protect\citeauthoryear{Cipollini and Gallo}{Cipollini and
  Gallo}{2022}]{Cipollini:Gallo:2022}
Cipollini, F. and G.~M. Gallo (2022, may).
\newblock Multiplicative error models: 20 years on.
\newblock {\em Econometrics and Statistics\/}.

\bibitem[\protect\citeauthoryear{Cipollini, Gallo, and Otranto}{Cipollini
  et~al.}{2021}]{Cipollini:Gallo:Otranto:2021}
Cipollini, F., G.~M. Gallo, and E.~Otranto (2021).
\newblock Realized volatility forecasting: Robustness to measurement errors.
\newblock {\em International Journal of Forecasting\/}~{\em 37\/}(1), 44 -- 57.

\bibitem[\protect\citeauthoryear{Cipollini, Gallo, and Palandri}{Cipollini
  et~al.}{2020}]{Cipollini:Gallo:Palandri:2020}
Cipollini, F., G.~M. Gallo, and A.~Palandri (2020).
\newblock Realized variance modeling: decoupling forecasting from estimation.
\newblock {\em Journal of Financial Econometrics\/}~{\em 18\/}(3), 532--555.

\bibitem[\protect\citeauthoryear{Cukierman}{Cukierman}{1986}]{Cukierman:1986}
Cukierman, A. (1986).
\newblock Central bank behavior and credibility: some recent theoretical
  developments.
\newblock {\em Federal Reserve Bank of St. Louis Review\/}~{\em 68\/}(5),
  5--17.

\bibitem[\protect\citeauthoryear{De~Luca and Zuccolotto}{De~Luca and
  Zuccolotto}{2011}]{DeLuca:Zuccolotto:2011}
De~Luca, G. and P.~Zuccolotto (2011).
\newblock A tail dependence-based dissimilarity measure for financial time
  series clustering.
\newblock {\em Advances in data analysis and classification\/}~{\em 5},
  323--340.

\bibitem[\protect\citeauthoryear{Dette, Golosnoy, and Kellermann}{Dette
  et~al.}{2022}]{Dette:Golosnoy:Kellermann:2022}
Dette, H., V.~Golosnoy, and J.~Kellermann (2022).
\newblock The effect of intraday periodicity on realized volatility measures.
\newblock {\em Metrika\/}, 1--28.

\bibitem[\protect\citeauthoryear{Engle}{Engle}{2002}]{Engle:2002}
Engle, R.~F. (2002).
\newblock New frontiers for {ARCH} models.
\newblock {\em Journal of Applied Econometrics\/}~{\em 17}, 425--446.

\bibitem[\protect\citeauthoryear{Engle and Gallo}{Engle and
  Gallo}{2006}]{Engle:Gallo:2006}
Engle, R.~F. and G.~M. Gallo (2006).
\newblock A multiple indicators model for volatility using intra-daily data.
\newblock {\em Journal of Econometrics\/}~{\em 131}, 3--27.

\bibitem[\protect\citeauthoryear{Engle and Lee}{Engle and
  Lee}{1999}]{Engle:Lee:1999}
Engle, R.~F. and G.~J. Lee (1999).
\newblock A permanent and transitory component model of stock return
  volatility.
\newblock In R.~F. Engle and H.~White (Eds.), {\em Cointegration, Causality,
  and Forecasting: A Festschrift in Honor of Clive W. J. Granger}, pp.\
  475--497. Oxford University Press, Oxford.

\bibitem[\protect\citeauthoryear{Engle and Russell}{Engle and
  Russell}{1998}]{Engle:Russell:1998}
Engle, R.~F. and J.~R. Russell (1998).
\newblock Autoregressive conditional duration: A new model for irregularly
  spaced transaction data.
\newblock {\em Econometrica\/}~{\em 66}, 1127--62.

\bibitem[\protect\citeauthoryear{Forsberg and Ghysels}{Forsberg and
  Ghysels}{2007}]{Forsberg:Ghysels:2007}
Forsberg, L. and E.~Ghysels (2007).
\newblock Why do absolute returns predict volatility so well?
\newblock {\em Journal of Financial Econometrics\/}~{\em 5}, 31--67.

\bibitem[\protect\citeauthoryear{Gallo, Lacava, and Otranto}{Gallo
  et~al.}{2021}]{Gallo:Lacava:Otranto:2021}
Gallo, G.~M., D.~Lacava, and E.~Otranto (2021).
\newblock On classifying the effects of policy announcements on volatility.
\newblock {\em International Journal of Approximate Reasoning\/}~{\em 134},
  23--33.

\bibitem[\protect\citeauthoryear{Hattori, Schrimpf, and Sushko}{Hattori
  et~al.}{2016}]{Hattori:Schrimpf:Sushko:2016}
Hattori, M., A.~Schrimpf, and V.~Sushko (2016, April).
\newblock The response of tail risk perceptions to unconventional monetary
  policy.
\newblock {\em American Economic Journal: Macroeconomics\/}~{\em 8\/}(2),
  111--36.

\bibitem[\protect\citeauthoryear{Huang and Tauchen}{Huang and
  Tauchen}{2005}]{Huang:Tauchen:2005}
Huang, X. and G.~Tauchen (2005).
\newblock The relative contribution of jumps to total price variance.
\newblock {\em Journal of financial econometrics\/}~{\em 3\/}(4), 456--499.

\bibitem[\protect\citeauthoryear{Hubert and Arabie}{Hubert and
  Arabie}{1985}]{Hubert:Arabie:1985}
Hubert, L. and P.~Arabie (1985).
\newblock Comparing partitions.
\newblock {\em Journal of classification\/}~{\em 2}, 193--218.

\bibitem[\protect\citeauthoryear{Johannes}{Johannes}{2004}]{Johannes:2004}
Johannes, M. (2004).
\newblock The statistical and economic role of jumps in continuous-time
  interest rate models.
\newblock {\em The Journal of Finance\/}~{\em 59\/}(1), 227--260.

\bibitem[\protect\citeauthoryear{Joyce, Lasaosa, Stevens, Tong, et~al.}{Joyce
  et~al.}{2011}]{Joyce:Lasaosa:Stevens:Tong:2011}
Joyce, M., A.~Lasaosa, I.~Stevens, M.~Tong, et~al. (2011).
\newblock The financial market impact of quantitative easing in the {UK}.
\newblock {\em International Journal of Central Banking\/}~{\em 7\/}(3),
  113--161.

\bibitem[\protect\citeauthoryear{Liao}{Liao}{2005}]{Liao:2005}
Liao, T.~W. (2005).
\newblock Clustering of time series data—a survey.
\newblock {\em Pattern recognition\/}~{\em 38\/}(11), 1857--1874.

\bibitem[\protect\citeauthoryear{Liu, Patton, and Sheppard}{Liu
  et~al.}{2015}]{Liu:Patton:Sheppard:2015}
Liu, L.~Y., A.~J. Patton, and K.~Sheppard (2015).
\newblock Does anything beat 5-minute {RV}? {A} comparison of realized measures
  across multiple asset classes.
\newblock {\em Journal of Econometrics\/}~{\em 187\/}(1), 293--311.

\bibitem[\protect\citeauthoryear{Maharaj, D'Urso, and Caiado}{Maharaj
  et~al.}{2019}]{Maharaj:DUrso:Caiado:2019}
Maharaj, E.~A., P.~D'Urso, and J.~Caiado (2019).
\newblock {\em Time series clustering and classification}.
\newblock Chapman and Hall/CRC.

\bibitem[\protect\citeauthoryear{McAleer and Medeiros}{McAleer and
  Medeiros}{2008}]{McAleer:Medeiros:2008}
McAleer, M. and M.~Medeiros (2008).
\newblock A multiple regime smooth transition heterogeneous autoregressive
  model for long memory and asymmetries.
\newblock {\em Journal of Econometrics\/}~{\em 147}, 104--119.

\bibitem[\protect\citeauthoryear{Otranto}{Otranto}{2008}]{Otranto:2008}
Otranto, E. (2008).
\newblock Clustering heteroskedastic time series by model-based procedures.
\newblock {\em Computational Statistics \& Data Analysis\/}~{\em 52\/}(10),
  4685--4698.

\bibitem[\protect\citeauthoryear{Otranto}{Otranto}{2015}]{Otranto:2015}
Otranto, E. (2015).
\newblock Capturing the spillover effect with multiplicative error models.
\newblock {\em Communications in Statistics-Theory and Methods\/}~{\em
  44\/}(15), 3173--3191.

\bibitem[\protect\citeauthoryear{Patton and Sheppard}{Patton and
  Sheppard}{2015}]{Patton:Sheppard:2015}
Patton, A.~J. and K.~Sheppard (2015).
\newblock Good volatility, bad volatility: Signed jumps and the persistence of
  volatility.
\newblock {\em Review of Economics and Statistics\/}~{\em 97\/}(3), 683--697.

\bibitem[\protect\citeauthoryear{Rand}{Rand}{1971}]{Rand:1971}
Rand, W.~M. (1971).
\newblock Objective criteria for the evaluation of clustering methods.
\newblock {\em Journal of the American Statistical association\/}~{\em
  66\/}(336), 846--850.

\bibitem[\protect\citeauthoryear{White}{White}{1980}]{White:1980}
White, H. (1980).
\newblock A heteroskedasticity-consistent covariance matrix estimator and a
  direct test for heteroskedasticity.
\newblock {\em Econometrica\/}~{\em 48\/}(4), 817--38.

\bibitem[\protect\citeauthoryear{Wright}{Wright}{2012}]{Wright:2012}
Wright, J.~H. (2012).
\newblock What does monetary policy do to long-term interest rates at the zero
  lower bound?
\newblock {\em The Economic Journal\/}~{\em 122\/}(564), F447--F466.

\end{thebibliography}
	
\end{document}